\documentclass[
	aps,
	showpacs,
	showkeys,
	twocolumn,
	altaffilletter,
	nolongbibliography,
	numerical,
	flushbottom,
	secnumarabic,
	pra,
	superscriptaddress,
	floatfix,
	10pt
]{revtex4-1}

\usepackage{lipsum}

\usepackage[dvipsnames]{xcolor}

\usepackage{bm}

\usepackage{graphicx}

\usepackage{upgreek}

\usepackage{natbib}

\usepackage{braket}

\usepackage{newtxtext, newtxmath}

\usepackage{siunitx}

\usepackage[inline]{enumitem}
\newlist{mylist}{enumerate*}{1}
\setlist[mylist]{label=\roman*)}

\usepackage{tikz}
\usetikzlibrary{arrows, external, intersections}
\usepackage{pgfplots, relsize}
\pgfplotsset{%
	compat = newest,%
	/pgf/number format/set thousands separator = {},%
	select coords between index/.style 2 args={
		x filter/.code={
			\ifnum\coordindex<#1\fi
			\ifnum\coordindex>#2\fi
		}
	}
}
\usepgfplotslibrary{groupplots, units, fillbetween}

\usepackage[caption=false]{subfig}
\usepackage[%
	breaklinks,%
	pdftex,%
	hyperfootnotes=true,%
	pdfpagelabels,%
	bookmarks,%
	pageanchor,%
]{hyperref}

\hypersetup{%
	colorlinks=true, linktocpage=true, pdfstartpage=1, pdfstartview=FitH, pdfborder={0 0 0},%
	breaklinks=true, pdfpagemode=UseNone, pageanchor=true, pdfpagemode=UseOutlines,%
	plainpages=false, bookmarksnumbered, bookmarksopen=true, bookmarksopenlevel=1,%
	hypertexnames=true, pdfhighlight=/O,
	urlcolor=blue, linkcolor=blue, citecolor=blue, 
}

\graphicspath{{Pictures/}}

\newcommand{\ie}{i.\,e.}

\newcommand{\eg}{e.\,g.}
\newcommand{\ped}[1]{_\text{#1}}
\newcommand{\api}[1]{^\text{#1}}
\newcommand{\nspin}{n}
\newcommand{\ham}{H}
\newcommand{\tf}{\tau}
\newcommand{\omegac}{\omega\ped{c}}
\newcommand{\spause}{s\ped{p}}
\newcommand{\lpause}{l\ped{p}}
\newcommand{\spauseopt}{s\ped{opt}}
\newcommand{\spausewor}{s\ped{wor}}
\newcommand{\sgap}{s_\Delta}
\newcommand{\diss}{\mathcal{D}}
\newcommand{\DELTA}{\mathop{}\!\Updelta}
\newcommand{\var}{\mathop{}\!\delta}
\newcommand{\rng}[2]{\ensuremath{[#1, #2]}}
\newcommand{\ev}[1]{\left\langle #1 \right\rangle}
\newcommand{\bigO}{O}
\renewcommand{\epsilon}{\varepsilon}
\newcommand{\eu}{e}
\newcommand{\iu}{i}

\begin{document}

\author{G.\,Passarelli}
\email{gpassarelli@fisica.unina.it}
\author{V.\,Cataudella}
\author{P.\,Lucignano}
\affiliation{Dipartimento di Fisica ``E.\,Pancini'', Universit\`a di Napoli ``Federico II'', Complesso di Monte S.~Angelo, via Cinthia, 80126 Napoli, Italy}
\affiliation{CNR-SPIN, c/o Complesso di Monte S. Angelo, via Cinthia - 80126 - Napoli, Italy}

\title{Improving quantum annealing of the ferromagnetic $ p $-spin model through pausing}

\date{\today}


\begin{abstract}
	The probability of success of quantum annealing can be improved significantly by pausing the annealer during its dynamics, exploiting thermal relaxation in a controlled fashion. In this paper, we investigate the effect of pausing the quantum annealing of the fully-connected ferromagnetic $ p $-spin model. This analytically solvable model has a search-like behavior and is often used as a benchmark for the performances of quantum annealing. We numerically show that \begin{mylist} \item the optimal pausing point is $ \SI{60}{\percent} $ longer than the avoided crossing time for the analyzed instance, and \item at the optimal pausing point, we register a \SI{45}{\percent} improvement in the probability of success with respect to a quantum annealing with no pauses of the same duration.\end{mylist} These results are in line with those observed experimentally for less connected models with the available quantum annealers. The observed improvement for the $ p $-spin model can be up to two orders of magnitude with respect to an isolated quantum dynamics of the same duration. 
\end{abstract}

\maketitle

\section{Introduction}\label{sec:intro}

Adiabatic quantum computation (AQC)~\cite{farhi:quantum-computation, albash:review-aqc, santoro-martonak} can be realized implementing quantum annealing (QA) algorithms~\cite{kadowaki:qa, dickson:thermal-qa} on quantum devices operating at low temperatures~\cite{harris:proto-dwave,harris:d-wave}. It is considered a heuristic technique for solving NP-hard optimization tasks that classical digital computers cannot solve efficiently. Defining and detecting a quantum speed up is a long-standing and delicate issue~\cite{ronnow:speedup, amin:freeze-out}, however it is known that quantum annealing, in some cases, performs better than its classical counterpart, thermal annealing~\cite{santoro-martonak, brooke:qa, albash:scaling-advantage}.

The main idea underlying AQC is to map the solution of a hard computational problem to the ground state of an Ising Hamiltonian~\cite{lucas:np-complete}. The two states of Ising spins represent the two logical states of classical bits.
At the starting time, the system is prepared in the ground state of a transverse field Hamiltonian, describing quantum spins fluctuating between their two classical values. These quantum fluctuations allow to explore the phase space through quantum tunneling. The Hamiltonian is slowly deformed to target the Ising Hamiltonian at the final time, hence the final ground state encodes the solution of the computational problem of interest. The adiabatic theorem of quantum mechanics ensures that a quantum  system, prepared in its ground state at time $ t = 0 $, will evolve remaining in its instantaneous ground state if the Hamiltonian is slowly varied in time, on a time scale proportional to the inverse of the minimal energy gap separating the ground state from the rest of the spectrum. In this framework, the time-to-solution is determined solely by the minimal energy gap.

However, in real annealing devices, mostly based on superconducting electronics~\cite{harris:proto-dwave, harris:d-wave}, such procedure runs in the presence of a finite-temperature environment, inducing dissipation. The current understanding of this problem is getting more and more detailed~\cite{amin:freeze-out, job:test-driving, marshall:thermalization}, and it is now clear that the time-to-solution is determined by a delicate balance between adiabaticity and thermal processes~\cite{amin:thermal-qa, mishra:finite-temperature-qa, albash:decoherence, lidar:open-systems, thorwart1, thorwart4, arceci:dissipative-lz, smelyanskiy:decoherence}.

In general, the environment is expected to be detrimental for quantum annealing~\cite{arceci:owp, keck:aqc}, despite the intrinsic robustness of AQC against decoherence~\cite{childs:robustness}. However, sometimes the thermal environment can improve quantum annealing performances~\cite{smelyanskiy:decoherence, passarelli:pspin, cangemi:beyond-born-markov}. This idea is interestingly pursued in a recent paper (Ref.~\cite{marshall}), in which the authors propose to pause the annealing at wisely chosen times, in order to take advantage also of thermal processes, in the search for the ground state of the final Hamiltonian. The possibility to pause the annealing dynamics is already implemented in the D-Wave 2000Q quantum computer~\cite{dwave-site}, and allowed the authors of Ref.~\cite{marshall} to experimentally test their hypothesis for instances of short range Ising models, easy to embed on the Chimera graph~\cite{choi:2008, choi:2011}. However, the recently announced Pegasus graph~\cite{pegasus}, with increased connectivity, can pave the way for embedding also $ p $-body all-to-all interacting models on the next generation of commercial quantum annealers, with less ancillary bits with respect to the Chimera graph~\cite{chancellor:max-k-sat, chancellor:multi-body-interactions}.

Indeed, computationally hard problems may require $ p $-body interactions (with $ p > 2 $), as for instance in Boolean satisfiability~\cite{cook:complexity} or in the Grover search~\cite{jorg:energy-gaps, grover:search}. Grover-like search can be described by the so-called fully-connected ferromagnetic $ p $-spin model (with large and odd $ p $), whose embedding on the Chimera graph can be cumbersome and inefficient (in particular for large values of $ p $), hence it is hardly studied with real devices. Nonetheless, this model has a remarkable relevance for many reasons. Originally introduced in Refs.~\cite{derrida:p-spin, gross:p-spin}, the $ p $-spin model is extensively studied in the context of spin-glasses~\cite{knysh:spin-glass, bapst:quantum-spin-glass, bapst:p-spin} and quantum optimization~\cite{seoane:transverse-interactions, nishimori:inhomogeneous-2, nishimori:reverse-pspin}. Its mean field character allows to recover analytic results in the thermodynamic limit for large $ p $. Moreover, the existence of an exact ground state, combined with the non-triviality of its phase diagram, makes it a natural candidate for benchmarking the performances of quantum annealing in an exactly solvable, realistically hard case~\cite{wauters:pspin}.

In this paper, we contribute to this topic investigating the quantum annealing of the fully-connected ferromagnetic $ p $-spin model. In particular, we focus on the possible advantages coming from pausing the annealing dynamics, as proposed in Ref.~\cite{marshall} for Ising-type models. This paper is organized as follows. In Section~\ref{sec:model}, we present the $ p $-spin model Hamiltonian and address the computation of the dynamics in the presence of a realistic dissipation model. We adopt a Born-Markov approximation and use a quantum master equation in the Lindblad form, simulated using a Monte Carlo wave function method~\cite{yip:mcwf}. In Section~\ref{sec:results}, we present our results, analyzing the effect of time and duration of the pause on the quantum annealing of a specific instance of the $ p $-spin model. Finally, we present our conclusions in Section~\ref{sec:conclusions}. We discuss the validity of the Born-Markov approximation in Appendix~\ref{app:lindblad}.

\section{Model}\label{sec:model}

\subsection{Ferromagnetic $ p $-spin model}\label{subsec:p-spin}

In this work, we adopt natural units and fix $ \hslash = k\ped{B} = 1 $. The dimensionless Hamiltonian of the $ p $-spin model for $ \nspin $ qubits reads
\begin{equation}\label{eq:pspin-hamiltonian}
	\ham\ped{p} = -\frac{\nspin}{2} {\left(\frac{1}{\nspin}\sum_{i = 1}^{\nspin} \sigma^z_i\right)}^p.
\end{equation}
For $ p $ odd, its ground state is ferromagnetic with all spins up-aligned. For $ p $ even, the system is $ Z_2 $-symmetric, and there are two degenerate ground states. The limit $ \nspin\to\infty $, $ p\to\infty $ (odd, with $ p \le \nspin $) of this model encodes a Grover-like  search~\cite{jorg:energy-gaps, grover:search}.

\begin{figure}[tb]
	\includegraphics[width = \linewidth]{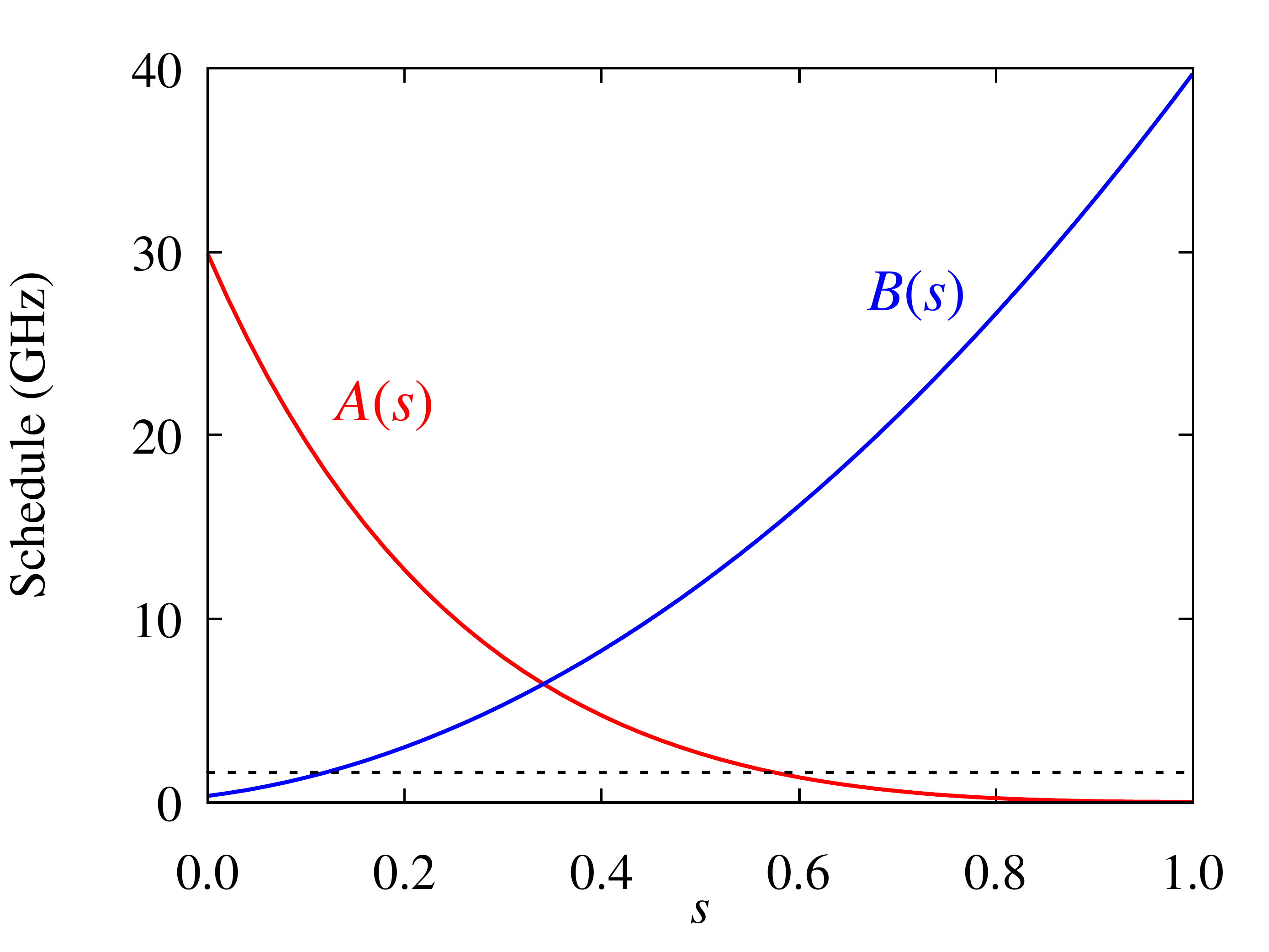}
	\caption{Annealing schedules of the D-Wave 2000Q. The dashed line is the working temperature of the device, equal to $ T = \SI{12.1}{\milli\kelvin} = \SI{1.57}{\giga\hertz} $.}
	\label{fig:schedules}
\end{figure}

The Hamiltonian of Eq.~\eqref{eq:pspin-hamiltonian} is reached via a quantum annealing starting from the dimensionless transverse field Hamiltonian $ \ham\ped{0} $:
\begin{equation}\label{eq:transverse-field}
	\ham_0 = -\frac{1}{2} \sum_{i = 1}^{\nspin} \sigma_x.
\end{equation}
We build the time-dependent Hamiltonian
\begin{equation}\label{eq:annealing-hamiltonian}
	\ham_Q(s) = A(s) \ham_0 + B(s) \ham\ped{p},
\end{equation}
where $ s = t/\tf \in \rng{0}{1} $ is the dimensionless time and $ \tf $ is the annealing time. $ A(s) $ and $ B(s) $ encode typical annealing schedules~\cite{job:test-driving, dwave-site}, and are depicted in Fig.~\ref{fig:schedules}. $ \ham_Q(s) $ is spherically symmetric, and both the ground state of $ \ham_0 $ and the ground state of $ \ham\ped{p} $ belong to the symmetry subspace of maximum spin, having dimension $ N = \nspin + 1 $.

\begin{figure}[b]
	\centering
	\includegraphics[width = \linewidth]{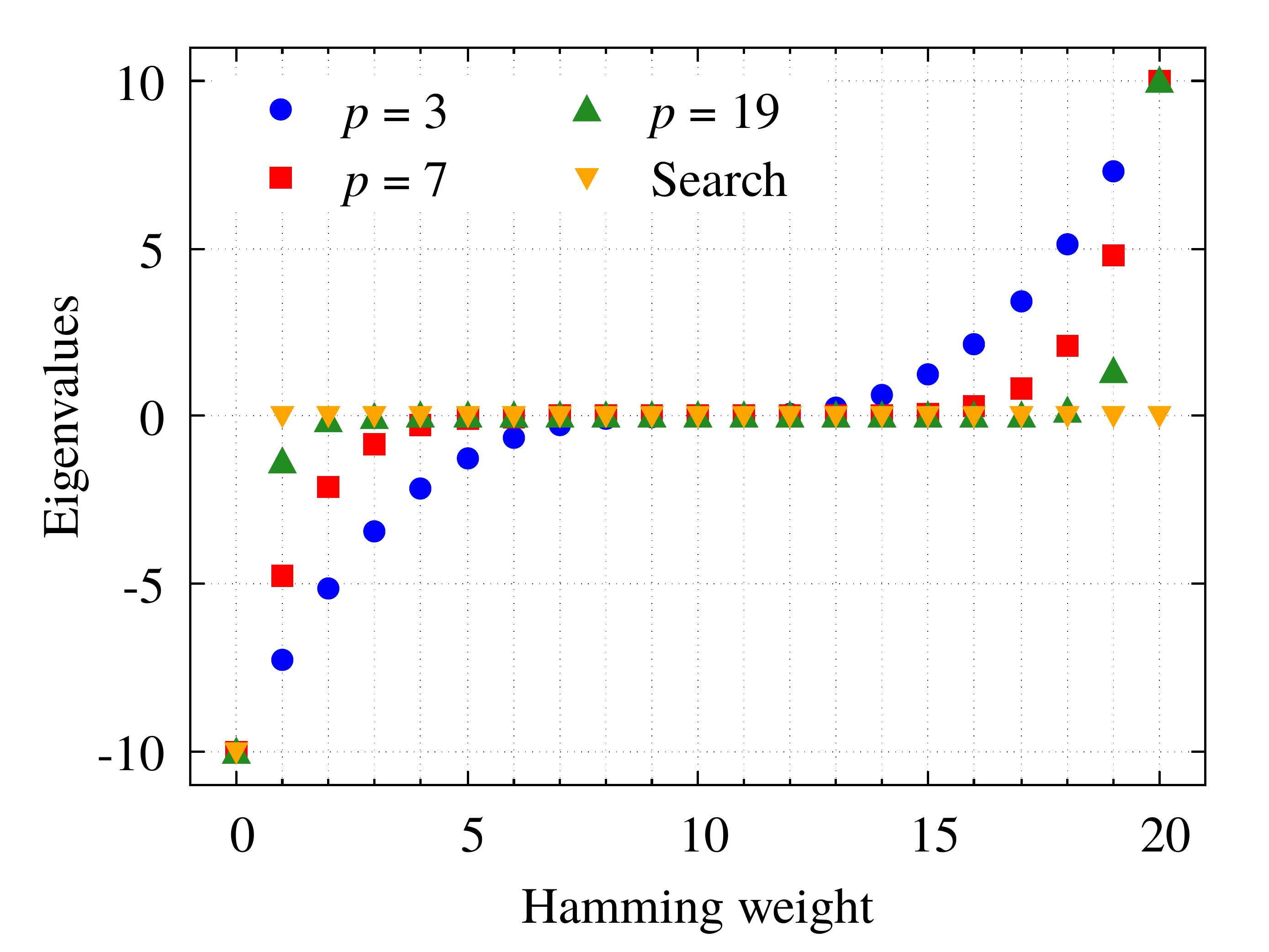}
	\caption{Eigenvalues of $ \ham\ped{p} $ [Eq.~\eqref{eq:pspin-hamiltonian}, dimensionless] versus the Hamming weight of the corresponding eigenstate, \ie, the number of qubits found in the state $ -1 $, for $ p = 3 $ (blue circle), $ p = 7 $ (red square), and $ p = 19 $ (green up triangle). Orange down triangles are the eigenvalues of $ \ham\ped{S} $ [Eq.~\eqref{eq:search}, dimensionless]. For $ p = 19 $, the $ p $-spin model shows features similar to the search task.}
	\label{fig:hammings}
\end{figure}

In order to obtain a quantum advantage for this adiabatic Grover-like search, the annealing schedule should be chosen carefully. Indeed, it is known that, using a linear schedule, the time to solution for search-like problems scales as $ 2^\nspin $, which is the same as in classical approaches~\cite{roland:adiabatic-search}. D-Wave's annealing schedule is not linear, but still qualitatively  different from the optimal annealing schedule for these classes of problems, reported in Refs.~\cite{roland:adiabatic-search, morley:search} (where the same quadratic speedup of the Grover~\cite{grover:search} algorithm is shown).

We fix the Hamiltonian parameters $ \nspin = 20 $ and $ p = 19 $ such that the $ p $-spin model Hamiltonian has a low lying spectrum very close to the one of the search Hamiltonian $\ham\ped{S}$:
\begin{equation}\label{eq:search}
	\ham\ped{S} = -\frac{\nspin}{2}\ket{0}\bra{0}.
\end{equation}
where $ \ket{0} $ is the ferromagnetic ground state with all spins up-aligned.
The comparison between the $ p $-spin and the search Hamiltonian is shown in Fig.~\ref{fig:hammings}, where the eigenvalues of $ \ham\ped{p} $ and $\ham\ped{S}$ are compared, as a function of the Hamming weight. The Hamming weight of a state is the number of $ -1 $ in its wave function, expressed in the single qubit basis $ \set{1, -1} $, and is a measure of the distance of the given state from the ferromagnetic ground state (whose Hamming weight is of course zero).
In Fig.~\ref{fig:hammings}, we compare the Hamming weights of the $ p $-spin model with $ p = 3 $, $ p = 7 $, and $ p = 19 $ with those of the search Hamiltonian of Eq.~\eqref{eq:search}.
The blue circles are for $ p = 3 $, the red squares are for $ p = 7 $, the green up triangles are for $ p = 19 $, and the orange down triangles are for the search Hamiltonian of Eq.~\eqref{eq:search}. Case $ p = 19 $ and search show very similar spectral properties. In fact, the eigenvalues of the $ p $-spin model with $ p = 19 $ are nonzero only close to the target state (and to the very excited state with opposite magnetization). By contrast, $ p = 3 $ and $ p = 7 $ show qualitatively different behaviors. 

In the thermodynamic limit $ \nspin\to\infty $, the $ p $-spin model is subject to a dynamical quantum phase transition (QPT)~\cite{heyl:dynamical-qpt}, at zero temperature, separating the para- and the ferromagnetic phases. The QPT occurs at $ s = \sgap $, \ie, the time of the avoided crossing between the two lowest-lying energy levels. The QPT is second-order for $ p = 2 $ and first-order for $ p > 2 $~\cite{bapst:quantum-spin-glass}. The latter reduces the efficiency of adiabatic computations as the minimal gap $ \Delta $ closes exponentially as a function of $ \nspin $~\cite{bapst:p-spin}. Several techniques are known to mitigate the detrimental effects of first-order QPTs in the quantum annealing of the $ p $-spin model, such as non-stoquastic AQC~\cite{seki:non-stoq, seoane:transverse-interactions, nishimori:non-stoq-2, nishimori:non-stoq}, inhomogeneous driving~\cite{nishimori:inhomogeneous-1, nishimori:inhomogeneous-2}, or reverse annealing~\cite{chancellor:reverse, perdomo:reverse, nishimori:reverse-pspin, dwave-site, d-wave-topology}. We will not discuss these techniques here.


\subsection{Environment and dissipation}

We model the environment as a collection of independent harmonic oscillators with Hamiltonian $ \ham_B = \sum_k \omega_k a_k^\dagger a_k $, linearly coupled to the qubit system via the interaction potential
\begin{equation}\label{eq:system-bath-hamiltonian}
	V\ped{QB} = g \sum_{i = 1}^\nspin \sigma_i^z \otimes \sum_k \left(a_k + a_k^\dagger\right),
\end{equation}
where $ g $ is the qubit-bath coupling energy and $ a_k $ ($ a_k^\dagger $) annihilates (creates) a boson in mode $ k $~\cite{caldeira:caldeira-leggett, Leggett}. We purposely use this operator as it does not break the spin symmetry, in contrast with more realistic dephasing models, where $ g $ becomes $ g_{ik} $. The annealer-environment Hamiltonian is $ \ham(t) = \ham_Q(t) + \ham_B + V\ped{QB} $.

Noise in superconducting quantum annealers is modeled as the sum of a low frequency $ 1/f $ contribution and a high frequency Ohmic spectrum~\cite{dwave-site, amin:decoherence, boixo:d-wave-noise, massarotti:breakdown}, having the form
\begin{equation}\label{eq:noise-ohmic}
	\gamma(\omega) = 2\uppi \eta \frac{\omega \eu^{-\omega / \omegac}}{1 - \eu^{-\beta \omega}},
\end{equation}
where $ \omegac $ is a high frequency cutoff, $ \beta = 1 / T $ and $ \eta $ is a dimensionless coupling defined via
\begin{equation}\label{eq:eta-definition}
	g^2 \sum_k \delta(\omega - \omega_k) = \eta \omega \eu^{-\omega / \omegac}.
\end{equation}
The spectral function in Eq.~\eqref{eq:noise-ohmic} satisfies the quantum detailed balance (\ie, the Kubo-Martin-Schwinger) condition~\cite{zanardi:master-equations}. We will neglect the $ 1 / f $ noise and fix $ \omegac = \SI{1}{\tera\hertz} $, $ T = \SI{12.1}{\milli\kelvin} = \SI{1.57}{\giga\hertz} $, and $ \eta = \num{1e-3} $. With these parameters, the annealer-bath coupling strength is weak compared with the energy scales of the device, as shown in Appendix~\ref{app:lindblad}. Thus, we can trace over bath states and adopt a Born-Markov approximation for the dynamics of the reduced system $ \rho(t) $~\cite{breuer:open-quantum}.

We employ a time-dependent Markovian quantum master equation (QME) in the Lindblad form to ensure complete positivity~\cite{lindblad76, gorini76, zanardi:master-equations}. It reads
\begin{equation}\label{eq:lindblad}
	\frac{d\rho(t)}{dt} = \iu \bigl[\rho(t), \ham_Q(t) + \ham\ped{LS}(t)\bigr] + \diss\bigl[\rho(t)\bigr],
\end{equation}
where the explicit form of the Lamb shift $ \ham\ped{LS} $ and the dissipator super-operator $ \diss $ is given in Appendix~\ref{app:lindblad}. They are both expressed in terms of Lindblad operators,
\begin{equation}\label{eq:lindblad-operators}
	L_{ab}(t) = \braket{\epsilon_a(t)| {\textstyle\sum_{k} \sigma_z^k} |\epsilon_b(t)} \ket{\epsilon_a(t)}\bra{\epsilon_b(t)},
\end{equation}
where $ \ket{\epsilon_a(t)} $ are the instantaneous eigenvectors of $ \ham_Q(t) $. Each Lindblad operator $ L_{ab} $ induces a quantum jump of frequency $ \omega_{ba}(t) = \epsilon_b(t) - \epsilon_a(t) $, from $ \ket{\epsilon_b(t)} $ to $ \ket{\epsilon_a(t)} $, for $ a \neq b $. Lindblad operators with $ a = b $ induce dephasing.

\subsection{Monte Carlo wave function}

Time-dependent Lindblad equations can be reformulated as stochastic Schr\"odinger equations for the qubit wave function~\cite{yip:mcwf}. This approach is similar to the quantum jump method in quantum optics~\cite{molmer:mcwf}. The wave function is evolved non-unitarily using an effective non-Hermitian Hamiltonian, which causes a loss of probability for the reduced system. The Lindblad operators make the system jump instantaneously between pairs of energy eigenstates, at random times and with rate $ \gamma(\omega) $, thus restoring the lost probability. The QME is recovered by taking averages over a number $ M $ of different stochastic realizations (or trajectories), with error scaling as $ M^{-1/2} $. The advantage of this method, known as Monte Carlo wave function (MCWF), is twofold. On the one hand, we work with ket states rather than density matrices, hence our software is less memory demanding. On the other hand, the trajectories are independent, thus the technique is easily parallelizable. In what follows, we fix $ M = 5000 $ trajectories as this number provides good accuracy in computations, \eg, a relative error $ \var{\ev{O}} / \ev{O} \sim \SI{1}{\percent} $ on all measured observables $ O $.

In order to unravel the QME through MCWF, Eq.~\eqref{eq:lindblad} has to be rewritten in a more suitable form. We follow the algorithm described in Ref.~\cite{yip:mcwf}. We collect all dephasing operators into a single one:
\begin{equation}\label{eq:dephasing}
L_0(t) = \sum_a \braket{\epsilon_a(t) | {\textstyle \sum_k \sigma_z^k} | \epsilon_a(t)} \ket{\epsilon_a(t)}\bra{\epsilon_a(t)}.
\end{equation}
Then, neglecting accidental degeneracies, we denote each non-zero frequency $ \omega_{ba} $ as $ \omega_\alpha $, with $ \alpha = 1, \dots, \nspin(\nspin - 1) $.
We label the corresponding Lindblad operators $ L_{ab}(t) $ (with $ a \ne b $) using the same index $ \alpha $. Defining also  $ \omega_{\alpha = 0} = 0 $, we now have $ \nspin (\nspin - 1) + 1 $ Lindblad operators, each uniquely accompanied by its frequency $ \omega_\alpha $, with $ \alpha = 0, \dots, \nspin(\nspin - 1) $.

Defining $ C_\alpha(t) = \sqrt{\gamma(\omega_\alpha)} L_\alpha(t) $, the dissipator in Eq.~\eqref{eq:lindblad} can be written compactly as
\begin{equation}
\diss\bigl[\rho(t)\bigr] = \sum_{\alpha} \Bigl[C_\alpha(t) \rho(t) C_\alpha^\dagger(t) - \frac{1}{2} \left\lbrace C^\dagger_{\alpha}(t) C_{\alpha}(t), \rho(t)\right\rbrace\Bigr].
\end{equation}
We rewrite Eq.~\eqref{eq:lindblad} as
\begin{align}\label{eq:lindblad-non-hermitian}
\frac{d \rho(t)}{dt} &= -\iu\left[\ham\ped{eff}(t)\rho(t) - \rho(t)\ham\ped{eff}^\dagger(t)\right] \notag \\
&\quad {}+ \sum_{\alpha} C_\alpha(t) \rho(t) C_\alpha^\dagger(t),
\end{align}
where the non-Hermitian Hamiltonian $ \ham\ped{eff}(t) $ reads
\begin{equation}\label{eq:effective-hamiltonian}
\ham\ped{eff}(t) = \ham_Q(t) + \ham\ped{LS}(t) - \frac{\iu}{2} \sum_{\alpha} C_\alpha^\dagger(t) C_\alpha(t).
\end{equation}
The non-Hermitian Hamiltonian of Eq.~\eqref{eq:effective-hamiltonian} generates a non-unitary dynamics. Given a starting state $ \ket{\psi(0)} $, we discretize the time interval using a time step $ \delta t $ and evolve the ket state using a first-order Trotter decomposition of the evolution operator,
\begin{equation}
	\ket{\tilde{\psi}(t)} = U(t, 0) \ket{\psi(0)} \approx \prod_{k \ge 0} \eu^{-\iu \ham\ped{eff}(k \delta t + \delta t / 2) \delta t} \ket{\psi(0)},
\end{equation}
where states with a tilde are not normalized. In fact, the environment shifts the energy levels of the spectrum of the Hamiltonian $ \ham_Q(t) $ and causes a decay of the norm of the state of the reduced system.
To ensure convergence, we progressively reduce $ \delta t $ until all analyzed observables are unaffected by the choice within Monte Carlo errors at all times. For an observable $ O(t) $, the MCWF error $ \sigma\ped{MC}(t) $ is defined as
\begin{equation}\label{eq:mc-errors}
	\sigma\ped{MC}^2(t) = \frac{1}{M (M - 1)} \sum_{m = 1}^{M} {\left[O_m (t) - \bar{O}(t)\right]}^2,
\end{equation}
where $ O_m(t) $ is the value assumed by $ O(t) $ in the $ m $th trajectory and $ \bar{O}(t) $ is the mean value over all trajectories.


Our numerical MCWF is implemented in Fortran 90 and employs waiting time distributions and the integrated algorithm described in Ref.~\cite{yip:mcwf},  computationally faster than step-by-step stochastic evolutions. We report here the pseudo code for the single quantum trajectory; the reader can find more details in the original reference.
\begin{enumerate}
	\item Draw a random number $ r $ uniformly distributed in $ \rng{0}{1} $.
	\item Starting from a normalized state $ \ket{\psi(0)} $, evolve non-unitarily using $ \ham\ped{eff}(t) $ until $ \braket{\tilde{\psi}(t^*) |\tilde{\psi}(t^*)} = r $. At $ t = t^* $, a quantum jump occurs. In fact, the norm squared of the unnormalized state $ \ket{\tilde{\psi}(t + \DELTA t)} $ at a time $ t + \DELTA t $ is the cumulative probability that no jumps have occurred during the time interval $ \rng{t}{t + \DELTA t} $:
	\begin{align}\label{eq:no-jumps-probability}
	p_0 &= 1 - \exp\left(-\int_{t}^{t + \DELTA t} \lambda(t') \, dt'\right) \notag\\
	&\equiv 1 - \braket{\tilde{\psi}(t + \DELTA t) | \tilde{\psi}(t + \DELTA t)},
	\end{align}
	where $ \lambda(t) $ is the jump rate and $ \DELTA t $ needs not to be infinitesimal. 
	\item Draw another random number $ \mu $ uniformly in $ \rng{0}{1} $ and select the quantum jump. The probability of jump $ \alpha $ is given by $ P_\alpha = \Pi_\alpha / \sum_{\alpha'} \Pi_{\alpha'} $, where $ \Pi_\alpha = \braket{\tilde{\psi}(t^*) | C_\alpha^\dagger C_\alpha | \tilde{\psi}(t^*)} $. The index of the occurring jump is the smallest non-negative integer $ m \le \nspin(\nspin - 1) $ satisfying $ \sum_{\alpha = 0}^{m} P_\alpha \ge \mu $.
	\item Update the state as $ \ket{\tilde{\psi}(t^* + 0^+)} = C_m \ket{\tilde{\psi}(t^*)} $ and renormalize. Draw another random number $ r $ and use $ \ket{\psi(t^* + 0^+)} $ as the new normalized starting state. Repeat steps \numrange[range-phrase = --]{1}{4} until the wanted annealing time.
\end{enumerate}

\subsection{Pausing the quantum annealing}

A feature of the D-Wave 2000Q allows to pause the quantum annealing at user defined dimensionless time $ \spause $ for a time duration $ \lpause $. During the time interval $ \lpause $, the system evolves with time-independent Hamiltonian $ \ham_Q(\spause) $, subject to dissipation. As discussed in Ref.~\cite{marshall}, pausing has effect on an isolated quantum system only if the pause is inserted at the minimal gap. Instead, open quantum systems may benefit from pausing later on during the dynamics. The environment provides relaxation channels towards the ground state. For the analyzed instances, the authors found an enhancement in the fidelity $ \Phi $, \ie, the ground state occupation probability at $ s = 1 $, if $ \spause $ is close to $ \sgap $, with the maximum improvement being for $ \spause $ approximately \SIrange{10}{15}{\percent} of the total annealing time longer than $ \sgap $. The authors interpret this result as a thermal-induced fidelity gain, concentrated around $ \sgap $ as the relaxation rate is maximum in this region. Consistently with this interpretation, they register no perceptible effects far enough from the minimal gap, either because the tunneling amplitude is large (\eg, at early times) or because the dynamics is frozen out (\eg, at long times)~\cite{amin:freeze-out}. Our aim is to study the same effect for the $ p $-spin model, using the QME in Eq.~\eqref{eq:lindblad}, unraveled using MCWF.

\begin{figure}[tb]
	\includegraphics[width = \linewidth]{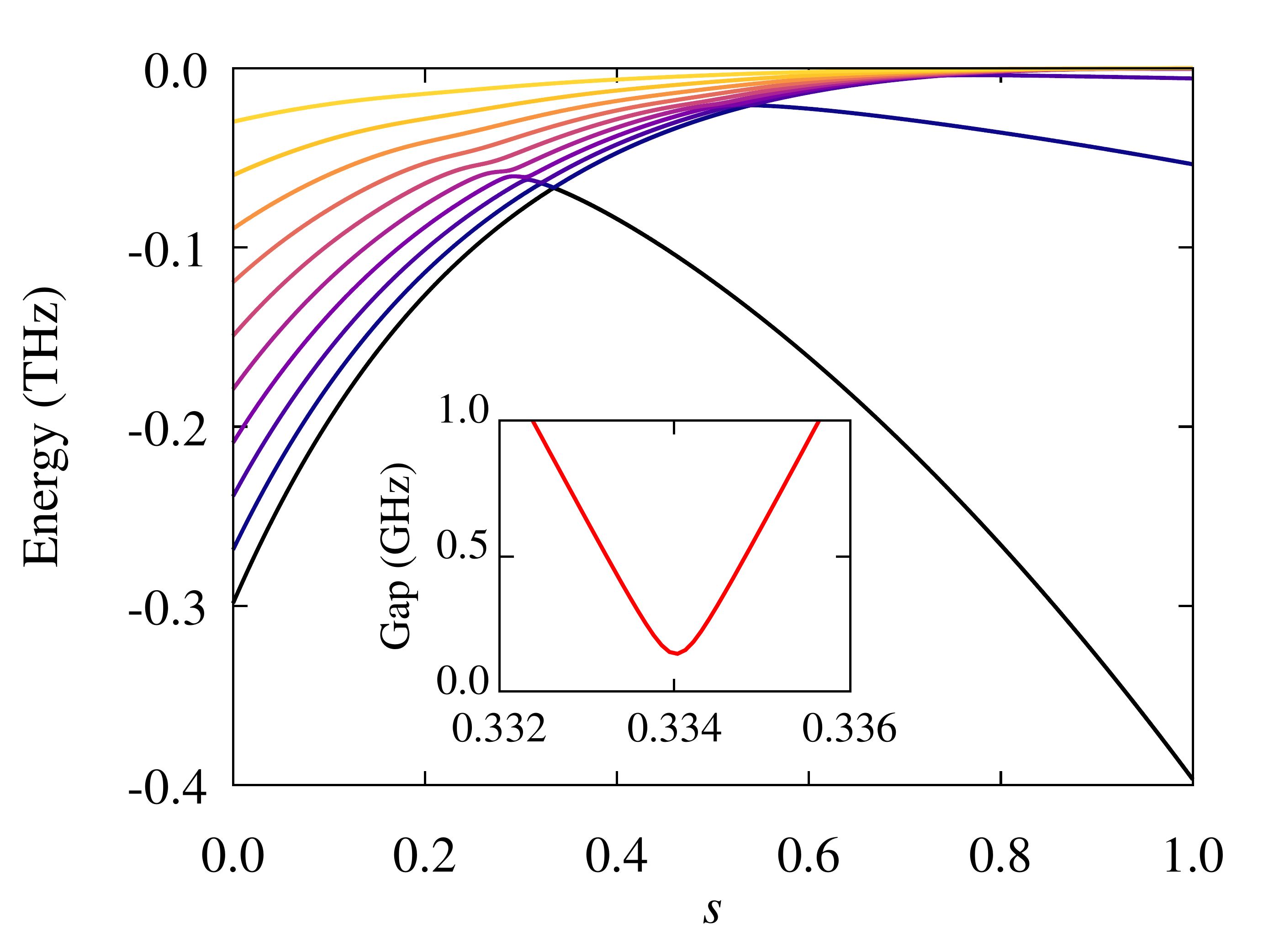}
	\caption{The $ L = 10 $ lowest-lying eigenvalues of the ferromagnetic $ p $-spin model with $ \nspin = 20 $ and $ p = 19 $, for our choice of parameters. The inset shows the instantaneous gap between the ground state and the first excited state, around $ s = \sgap \approx 0.334 $. Units are defined with $ \hslash = 1 $.}
	\label{fig:spectrum}
\end{figure}

We present our results in the next Section. In the literature, the dissipative dynamics of the $ p $-spin model has been studied for systems up to $ \nspin = 16 $ qubits and $ p \le 7 $~\cite{passarelli:pspin, passarelli:proceeding, nakamura:sboroni}. Here, we study a system of $ \nspin = 20 $ qubits, with $ p = 19 $.  With our parameters, the minimal gap is $ \Delta \approx \SI{0.14}{\giga\hertz} $, one order of magnitude less than the working temperature of the device. The avoided crossing between the two lowest energy levels occurs at $ \sgap \approx 0.334 $. In Fig.~\ref{fig:spectrum}, we report the first $ L = 10 $ eigenvalues and show the minimal gap $ \Delta $, obtained via numerical diagonalization of $ \ham_Q(s) $.

We choose the annealing time $ \tf = \SI{100}{\nano\second} $. This is one order of magnitude smaller than the minimal annealing time of currently available quantum annealers, and might be experimentally accessible with next-generation D-Wave devices~\cite{job:test-driving}. In the unitary limit, this choice of $ \tf $ makes the dynamics non-adiabatic, with a fidelity $ \Phi \approx \num{5.51e-3} $. Of course, one may decide to choose a longer annealing time to obtain a higher fidelity. For instance, with $ \tf' = 10 \tf $ one obtains $ \Phi \approx \num{5.39e-2} $, \ie, an improvement of one order of magnitude. However, in Refs.~\cite{passarelli:pspin, passarelli:proceeding}, the authors have shown that in this non-adiabatic regime the low temperature environment can improve the annealing performances. Moreover, pausing the system can further increase the fidelity within the same time scales, as shown in the next Section. 

\section{Results}\label{sec:results}

\subsection{Quantum annealing with no pauses}

\begin{figure}[tb]
	\centering
	\includegraphics[width = \linewidth]{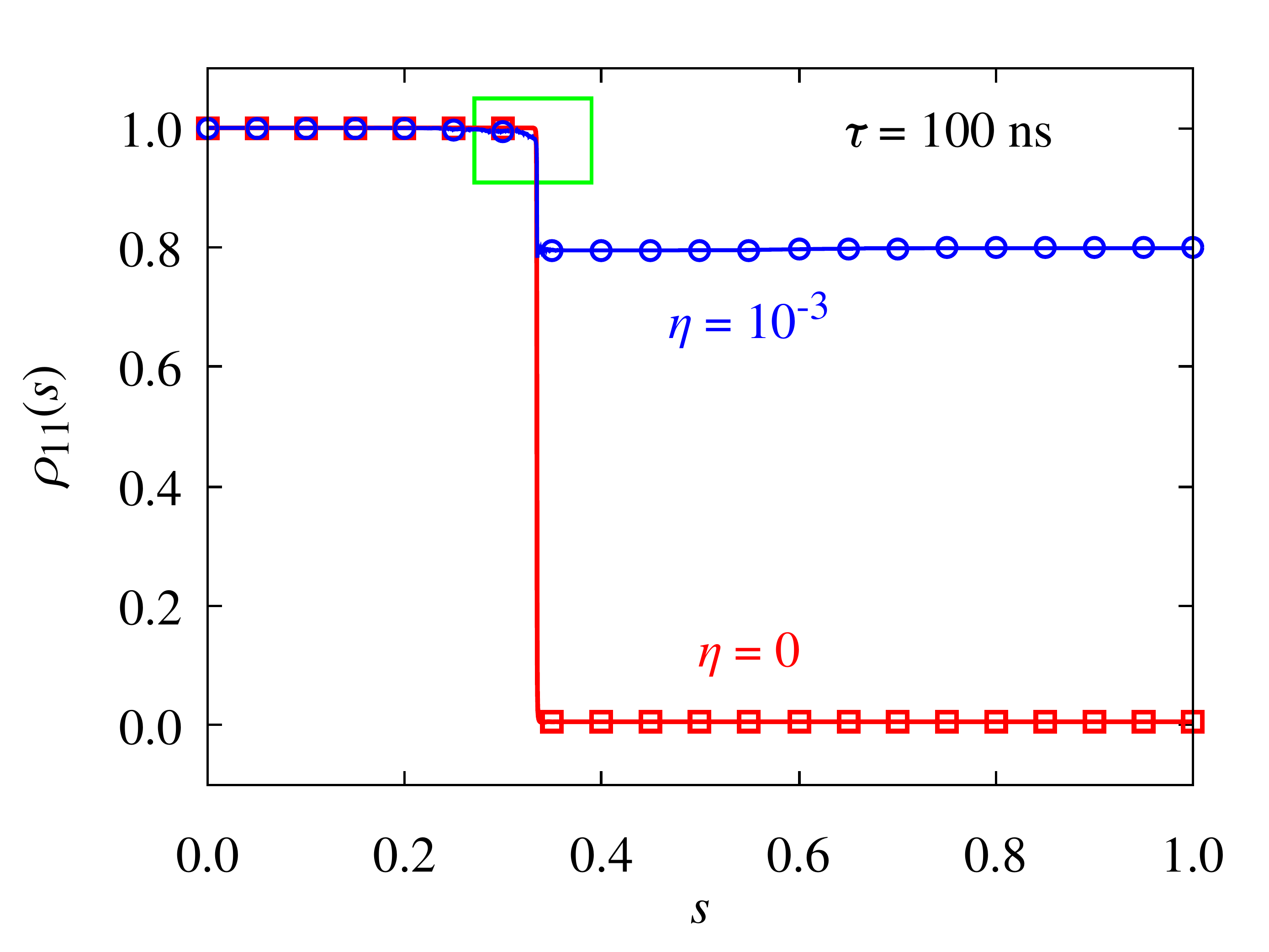}
	\caption{Ground state probability $ \rho_{11}(s) $ as a function of the dimensionless time $ s $, for unitary ($ \eta = 0 $) and dissipative dynamics ($ \eta = \num{1e-3} $), with $ \tf = \SI{100}{\nano\second} $. Other parameters are given in the main text. The dynamics is subject to Landau-Zener transitions, reducing the ground state occupation probability at $ \sgap \approx 0.334 $. This effect is softened by thermal decays in the presence of the environment. In the main text, we explain the decrease of $ \rho_{11}(s) $ in the dissipative case (\eg, the highlighted area in the picture).}
	\label{fig:dynamics}
\end{figure}

The annealing time $ \tf $ is smaller than $ h / \Delta^2 $, where $ h = \max_{s,a,b} \braket{\epsilon_a(s) | \partial_s \ham_Q(s) | \epsilon_b(s)} \approx \SI{1.2}{\tera\hertz} $, hence the quantum annealing is not adiabatic~\cite{zanardi:master-equations}. Landau-Zener diabatic transitions~\cite{landau:crossings, zener:crossings}  excite the qubit system and abruptly reduce the fidelity at the avoided crossing between the ground state and the first excited state. 
In Fig.~\ref{fig:dynamics}, we show the ground state population $ \rho_{11}(s) $ in the energy eigenbasis as a function of time, during a dynamics driven by the Hamiltonian in Eq.~\eqref{eq:annealing-hamiltonian}. The red line with squares represents the unitary dynamics ($ \eta = 0 $) and the blue line with circles the dissipative one ($ \eta = \num{1e-3} $). 

In the unitary case, the ground state population drops almost to zero after the avoided crossing at $ s = \sgap $. Far from $ \sgap $, the dynamics is frozen due to the large level spacing and the population remains constant. The fidelity of the algorithm is $ \Phi \approx \num{5.51e-3} $. 

In the dissipative case, after $ s = \sgap $, the bath-induced relaxation partially compensates diabatic transitions and improves the fidelity ($ \Phi \approx 0.799 $) with respect to the isolated case. Instead, bath-induced excitations right before the gap account for the small decrease in the plateau that can be seen for $ s \lesssim \sgap $. Far from $ s = \sgap $, thermal processes are exponentially suppressed and the ground state occupation probability is constant, in analogy with the isolated dynamics~\cite{marshall}.

\begin{figure*}[t]
	\centering
	\subfloat[]{\label{fig:pause-length}\includegraphics[width = 0.49\linewidth]{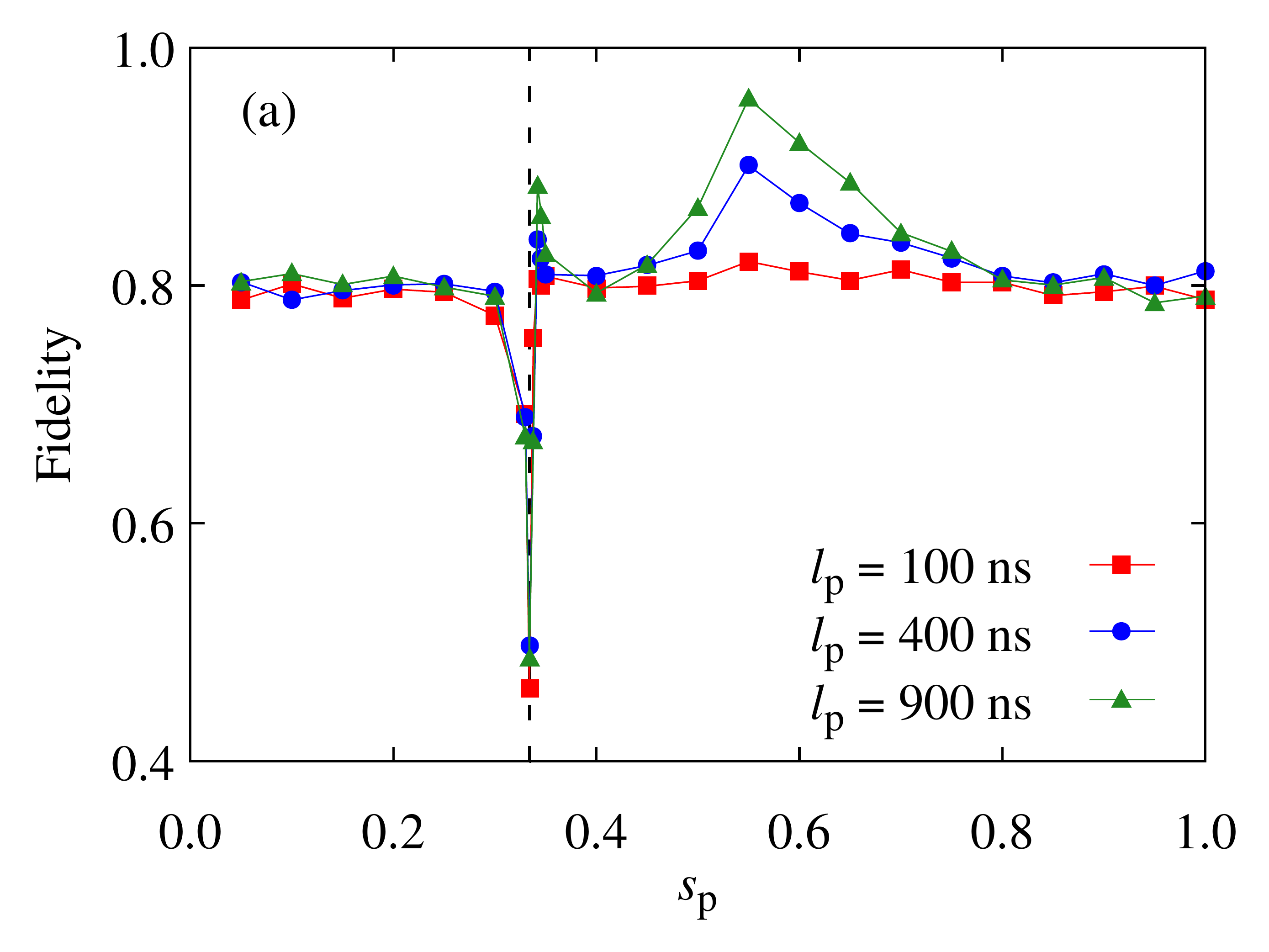}}\hfill
	\subfloat[]{\label{fig:pause-length-zoom}\includegraphics[width = 0.49\linewidth]{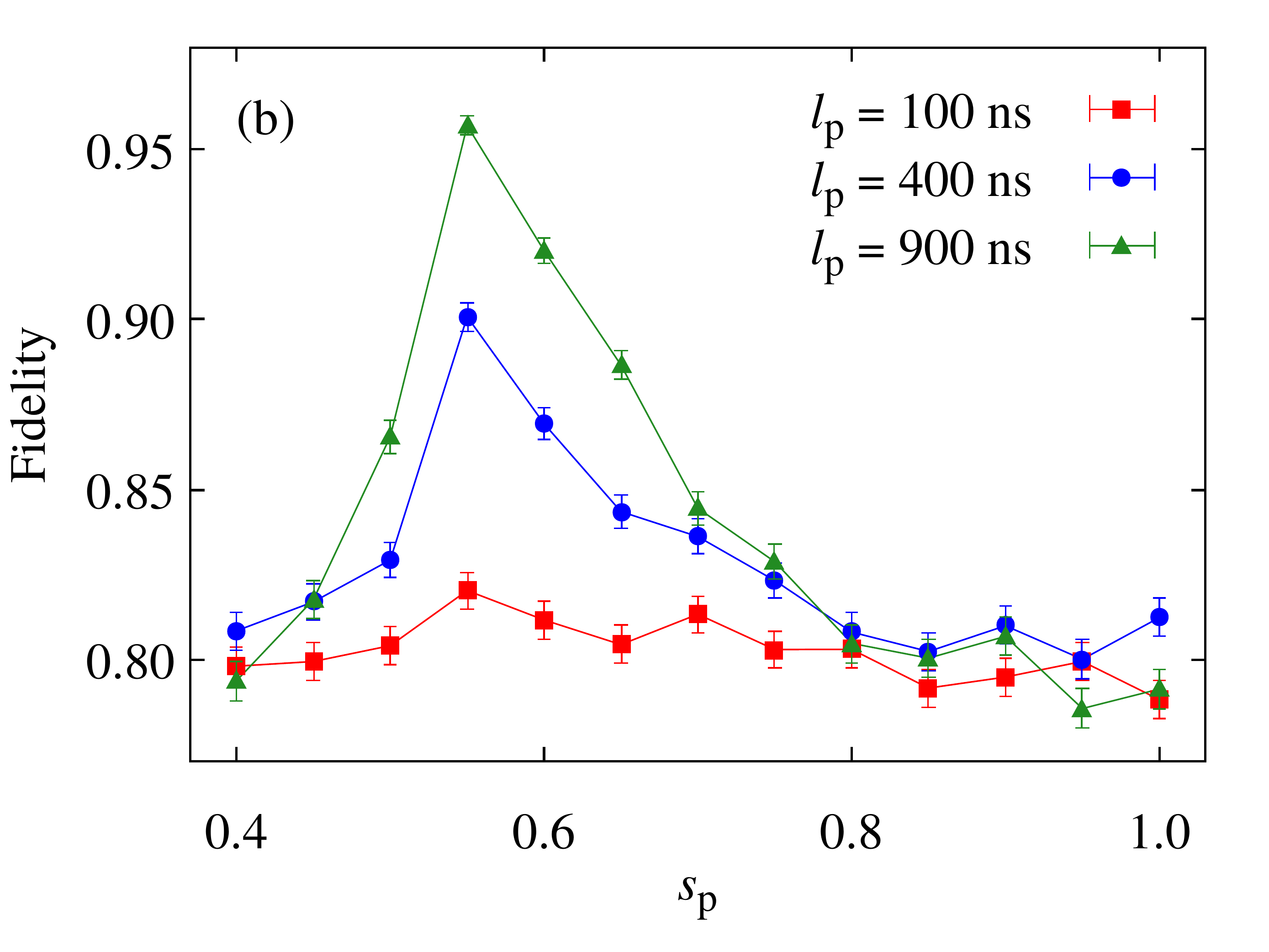}}
	\caption{Fidelity of the quantum annealing as a function of the pausing time $ \spause $, for several values of the pause length $ \lpause $. Error bars are the standard errors of MCWF [see Eq.~\eqref{eq:mc-errors}]. In Fig.~\ref{fig:pause-length}, error bars are of the order of the point size, and the dashed vertical line indicates the time of the minimal gap, $ \sgap \approx 0.334 $. In Fig.~\ref{fig:pause-length-zoom}, we focus on the region around the optimal pausing point $ \spauseopt \approx 0.55 $.}
	\label{fig:pause}
\end{figure*}

\begin{figure*}[t]
	\centering
	\subfloat[]{\label{fig:heatmap}\includegraphics[width = 0.475\linewidth]{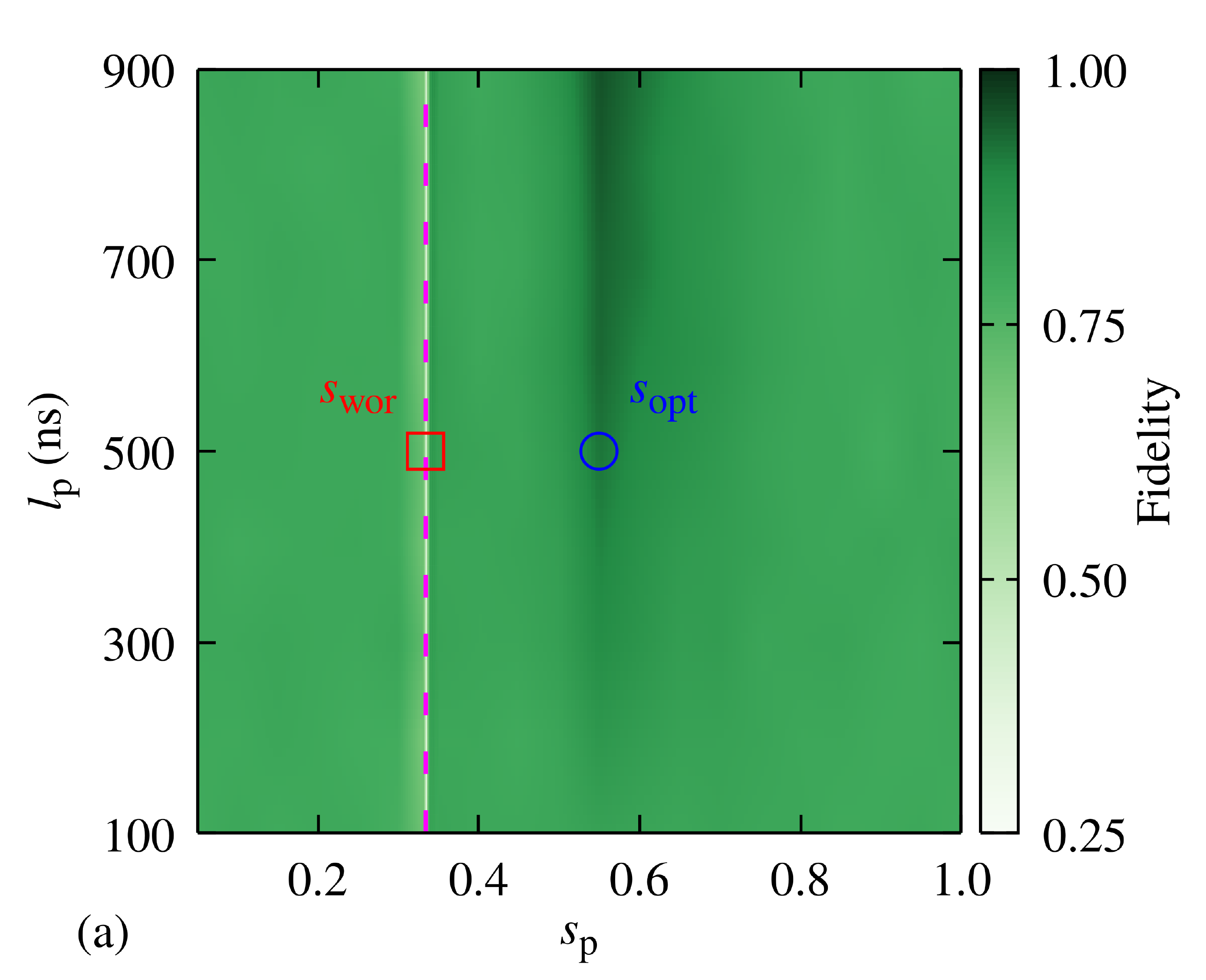}}\hfill
	\subfloat[]{\label{fig:heatmap-zoom}\includegraphics[width = 0.475\linewidth]{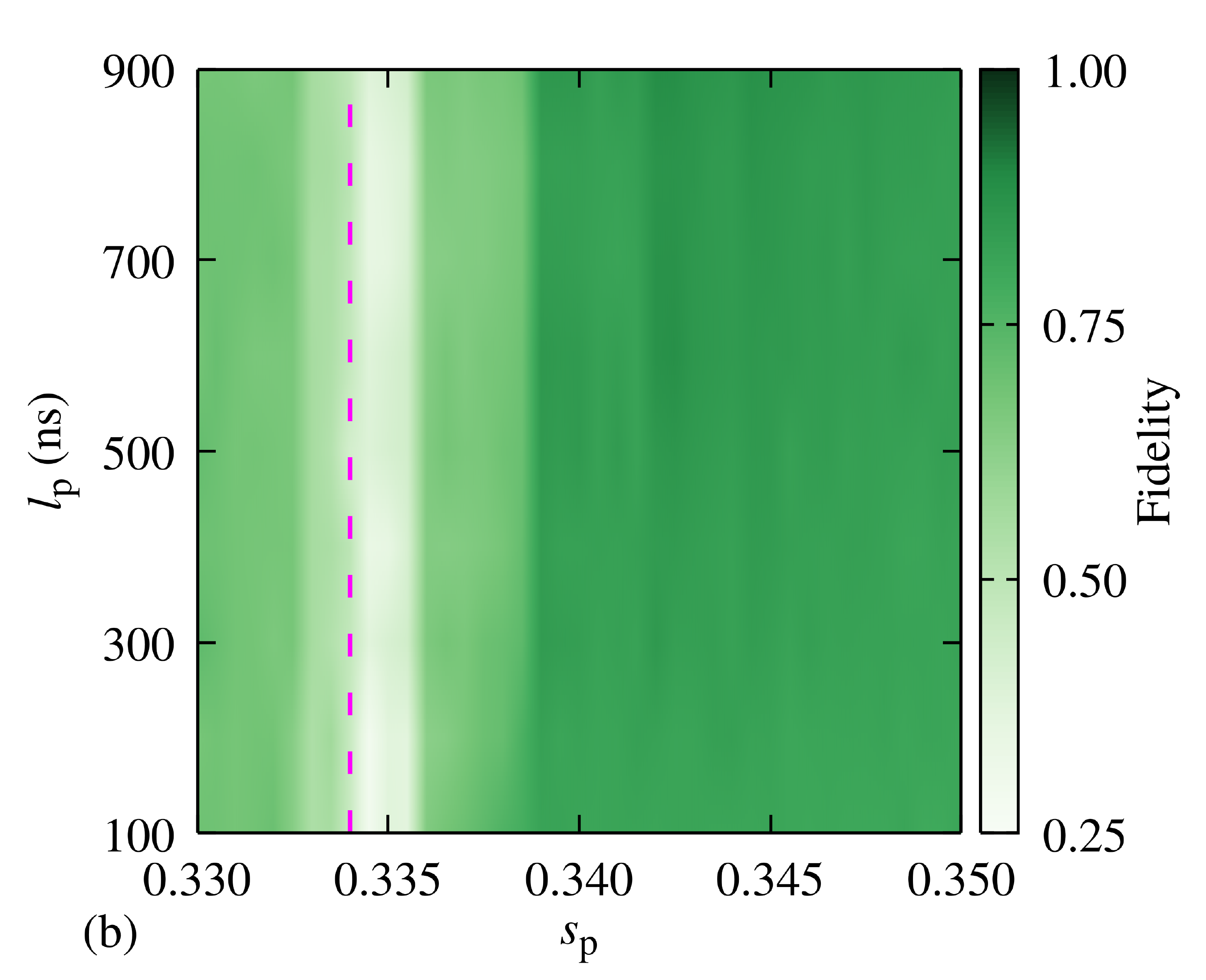}}
	\caption{Density plots of the fidelity as a function of pause time $ \spause $ and duration $ \lpause $, for $ \spause\in\rng{0}{1} $ and $ \lpause\in\rng{\SI{100}{\nano\second}}{\SI{900}{\nano\second}} $. The qualitative behavior is independent of $ \lpause $. The quantitative differences are more pronounced around $ \sgap $ and around the optimal pausing time $ \spauseopt \approx 0.55 $. In Fig.~\ref{fig:heatmap}, we marked with symbols the two points $ (\spause = \spausewor, \lpause = \SI{500}{\nano\second}) $ (red square) and $ (\spause = \spauseopt, \lpause = \SI{500}{\nano\second}) $ (blue circle), whose corresponding dynamics is shown in Fig.~\ref{fig:pause-time}. Fig.~\ref{fig:heatmap-zoom} is a focus around the gap. In both panels, the dashed magenta line marks the time of the minimal gap $ \sgap \approx 0.334 $.}
	\label{fig:density-plots}
\end{figure*}	

In what follows, we adopt a two-level model to give a simple description of the small decrease in $ \rho_{11}(s) $ before $ \sgap $, highlighted in Fig.~\ref{fig:dynamics}. Let us consider the ground state and the first excited state, and focus on the region $ s\in\rng{s_T}{\sgap} $, where $ s_T $ is the time before $ \sgap $ where the instantaneous gap is equal to the temperature $ T $. At shorter times, the thermal processes have scarce influence on the dynamics. At $ s = s_T $, the system is almost completely in its ground state. As our simplifying hypothesis, we suppose that for $ s\in\rng{s_T}{\sgap} $ the gap is constant and equal to $ \Delta $. In the energy eigenbasis, the density matrix of the qubit system at $ s = s_T $ is diagonal as $ B(s \le s_T) \ll A(s \le s_T) $, and can be written as
\begin{equation}
	\rho(s_T) = 
	\begin{pmatrix}
		\rho_{11}(s_T) & 0 \\
		0 & \rho_{22}(s_T) 
	\end{pmatrix},
\end{equation}
where $ \rho_{11}(s_T) \approx 1 $ and $ \rho_{22}(s_T) = 1 - \rho_{11}(s_T) \approx 0 $. The population transfer due to thermal processes can be effectively modeled  by a classical master equation of the form
\begin{equation}\label{eq:classical-me}
	\frac{1}{\tf}\frac{d\rho_{11}(s)}{ds} = \Gamma_{2\to1}\rho_{22}(s) - \Gamma_{1\to2} \rho_{11}(s), 
\end{equation}
where $ \Gamma_{1\to2} = \gamma(-\Delta) = \eu^{-\beta\Delta} \gamma(\Delta) $ and $ \Gamma_{2\to1} = \gamma(\Delta) $. The solution to this ME reads
\begin{equation}
	\rho_{11}(s) = \rho_{11}(s_T) - C \bigl[1 - \eu^{-(s - s_T) / s_1}\bigr],
\end{equation}
where $ C = \rho_{11}(s_T) - \tau s_1 \gamma(\Delta) $ and $ s_1 = 1 / \tau \sqrt{\gamma(\Delta)} \bigl(1 + \eu^{-\beta \Delta}\bigr) $. The ground state population at $ s \to \sgap^{-} $ is then $ \rho_{11}(\sgap^-) \approx 0.975 $, in agreement with numerical simulations.

\subsection{Quantum annealing with a pause}

Thermal effects can be enhanced by pausing the quantum annealing. We tested various pause lengths $ \lpause $, from \SIrange{100}{900}{\nano\second}. The maximum pause length $ \lpause $ gives a total annealing time $ \tf' = 10 \tf $, whose corresponding fidelity, in the absence of pauses, is $ \Phi \approx \num{5.39e-2} $ for the unitary dynamics and $ \Phi \approx 0.664 $ for the dissipative one, with $ \eta = \num{1e-3} $. The effects of pausing substantially increases this fidelity. These results are $ \lpause $-independent, although there are small quantitative differences. In Fig.~\ref{fig:pause-length}, we show the fidelity $ \Phi $ as a function of the pause time $ \spause \in \rng{0}{1} $, for different values of the pause duration, $ \lpause = \text{\SIlist{100; 400; 900}{\nano\second}} $. Fig.~\ref{fig:pause-length-zoom} is a zoom in the region around the optimal pausing time $ \spauseopt \approx 0.55 $. Four different regions can be distinguished for all $ \lpause $.
\begin{enumerate}
	\item When $ \spause < \sgap $, the fidelity is not affected by pausing. Here, the tunneling amplitude is large compared with the thermal relaxation rate, and the system evolves quantum mechanically with little influence from the environment.
	\item When $ \spause \approx \sgap $, thermal processes are more frequent and, correspondingly, the relaxation rate is maximum. In this region, $ \exp(-\beta \Delta) \sim 1 $, hence the excitation rate $ \Gamma_{1\to2} $ is comparable with the decay rate $ \Gamma_{2\to1} $. When $ \spause \lesssim \sgap $, most of the population is in the adiabatic ground state. Thus, transitions from the ground state to the first excited state are more probable than reverse processes. This imbalance causes a decrease in the observed fidelity after the pause, and the worst pausing time is sharply localized right before $ \sgap $. The situation is reversed for $ \spause \gtrsim \sgap $, and here the fidelity is slightly enhanced. This effect is more pronounced for longer $ \lpause $.
	\item For $ \spause = \spauseopt $, we observe a peak in the success probability for any $ \lpause > \SI{100}{\nano\second} $. The peak height increases with increasing $ \lpause $, following a saturation law of the form
	\begin{equation}\label{eq:saturation}
		\Phi(\lpause) = \Phi\ped{sat}\left[1 - \alpha \eu^{-(\lpause - l_0) / T\ped{r}}\right],
	\end{equation}
	with fitted parameters $ \Phi\ped{sat} = 0.976 \pm 0.007 $, $ \alpha = 0.160 \pm 0.005 $, and $ T\ped{r} = \SI{4.1}{\nano\second} \pm \SI{0.4}{\nano\second} $, and $ l_0 $ fixed to $ l_0 = \SI{100}{\nano\second} $. In particular, $ T\ped{r} $ is related to the thermal relaxation time of the many-body system, and $ \Phi\ped{sat} $ is an estimate of the maximum fidelity that can be achieved by pausing the dynamics at the optimal point. The fidelity shows a peak almost at the time $ \spause = \spauseopt $ independently of the pause duration $ \lpause $. By contrast, $ \lpause $ influences the time at which the fidelity goes back to its baseline value. In particular, for $ \lpause = \SI{900}{\nano\second} $, we register a $ \SI{20}{\percent} $ increase of the fidelity with respect to the dissipative dynamics with no pause and total annealing time $ \tf $, and a $ \SI{45}{\percent} $ increase with respect to a dissipative dynamics of total annealing time $ \tf' $. The optimal pausing point $ \spauseopt $ is independent of the pause length. 
	\item When $ \spause \gg \sgap $, the fidelity is not influenced much by pausing. The ground state is well-separated in energy from the other levels (see Fig.~\ref{fig:spectrum}).
	The eigenstates of the Hamiltonian are almost diagonal in the $ \sigma_z $ basis and the qubit-bath coupling operator has exponentially small off-diagonal matrix elements. Perturbation theory predicts that thermal processes are exponentially suppressed, and the dynamics is frozen.
\end{enumerate}	

All the results are summarized in Fig.~\ref{fig:density-plots}, where we show the fidelity as a function both of pause time and pause duration. Fig.~\ref{fig:heatmap-zoom} is a zoom in the region around $ \sgap $. The heat maps show even more evidently that around an optimal pausing time $ \spauseopt $ the fidelity abruptly increases, almost independently of $ \lpause $. This is evident in Fig.~\ref{fig:density-plots}, where the dark shadow shows up at $ \spause \approx \spauseopt $ and $ \lpause > \SI{100}{\nano\second} $. At shorter pausing lengths $ \lpause < \SI{100}{\nano\second} $, this phenomenon is no longer visible. The white vertical line around $ \spause \approx \sgap $ reflects a sharp decrease of the fidelity and can be better visualized in Fig.~\ref{fig:heatmap-zoom}. For the fully-connected $ p $-spin model, the largest fidelity enhancement occurs for $ \spauseopt \approx 1.65 \, \sgap $, or, equivalently, $ \spauseopt \approx \sgap + 0.22 $, while the worst pausing point is $ \spausewor \approx \sgap $. 

\begin{figure}[tb]
	\centering
	\includegraphics[width = \linewidth]{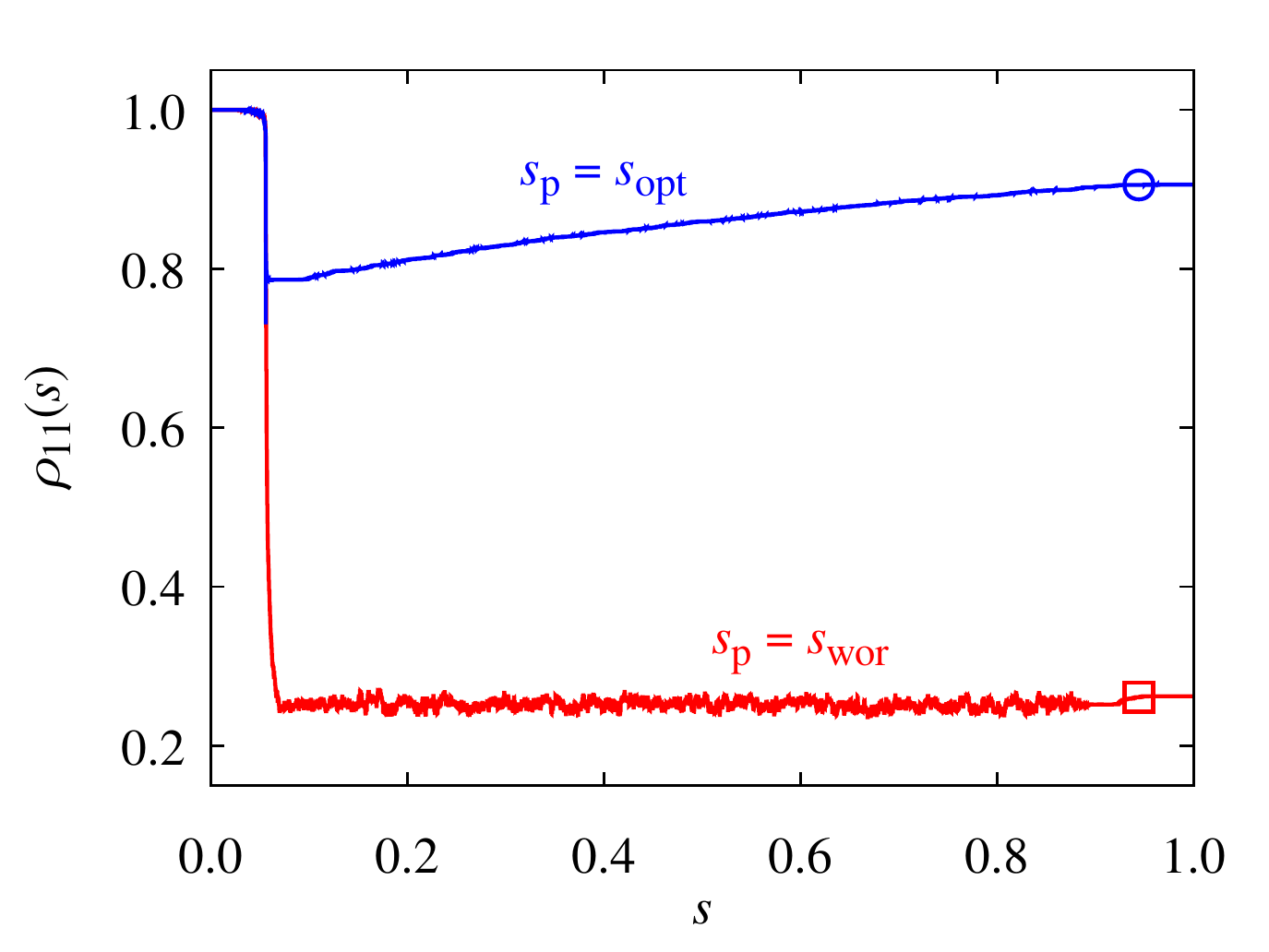}
	\caption{$ \rho_{11}(s) $ as a function of $ s $, with annealing time $ \tau = \SI{100}{\nano\second} $. A pause of length $ \lpause = \SI{500}{\nano\second} $ is inserted either at $ \spause = \spauseopt $ (blue curve with a circle) or at $ \spause = \spausewor $ (red curve with a square).}
	\label{fig:pause-time}
\end{figure}

Fig.~\ref{fig:pause-time} shows the differences in the dynamics of $ \rho_{11}(s) $ when a pause of $ \lpause = \SI{500}{\nano\second} $ is inserted at $ \spauseopt $ (blue curve with a circle) or $ \spausewor $ (red curve with a square). When $ \spause = \spauseopt $, the ground state population grows monotonically during the pause because of thermal relaxation. In fact, thermal excitations outside the ground state are suppressed as the spectral gap is large compared to $ T $, \eg, $ \epsilon_2(\spauseopt) - \epsilon_1(\spauseopt) \approx \SI{100}{\giga\hertz} $. On the other hand, for $ \spause = \sgap $, the system can be excited due to $ \Delta \ll T $ and $ \rho_{11}(s) $ is reduced with respect to the previous case. Excitations and decays are almost equally probable, and produce evident noisy oscillations of $ \rho_{11}(s) $ around a stationary value $ \bar{\rho}_{11} \approx 0.25 $ during the pause.

We repeated the simulations of this Section using Hamiltonian~\eqref{eq:search} as target Hamiltonian. Results are presented in Appendix~\ref{app:search}.

\section{Conclusions}\label{sec:conclusions}

Quantum annealing is designed for solving NP-hard optimization tasks faster than classical techniques. This class of problems can often be mapped onto Ising Hamiltonians, and solved heuristically using the available hardware. The last generation of D-Wave devices has made possible to fine tune the annealing schedule, through reverse annealing, quenches, qubit-specific annealing offsets, and pausing. These advanced strategies allow to study several interesting problems. Currently, one of the main limitations is the sparse connectivity of the Chimera graph, since in many cases a cost function involves long range and multiple-body interactions among qubits. These problems are hard to embed efficiently in modern quantum annealers, as they require a large number of ancilla qubits and minor or parity embedding~\cite{choi:2008, choi:2011, chancellor:multi-body-interactions}. In the context of coherent Ising machines~\cite{yamamoto:coherent-ising-machines}, there have already been several proposals of all-to-all architectures based on superconducting electronics~\cite{puri:all-to-all, nigg:full-connectivity} or quantum optical systems~\cite{inagaki:dopo, mcmahon:100-spin, inagaki:2000-nodes}, that should allow to overcome these limitations. At present, numerical simulations are fundamental to capture the physical behavior of fully-connected systems, while these promising platforms are developed and refined.


In this paper, we studied the quantum annealing of the ferromagnetic $ p $-spin model for $ \nspin = 20 $ and $ p = 19 $. The embedding of this model requires a large number of ancillae, for instance, using the decomposition of $ p $-body interactions into $ 2 $-body interactions described in Ref.~\cite{zoller:many-body-into-pairs}, or adopting the perturbative scheme described in Ref.~\cite{dodds:practical-designs}, with a number of auxiliary qubits equal to the number of logical qubits. Instead, due to the spin symmetry of this model, we are able to simulate its dissipative dynamics using MCWF. Using a realistic model of dissipation, we showed that at anti-adiabatic times quantum annealing can be improved in the presence of a low temperature environment. More importantly, the beneficial thermal effects can be enhanced by pausing the annealing around an optimal pausing point. In fact, we showed that a quantum annealing of total time $ \tf' = \SI{1000}{\nano\second} $ is circa \SI{45}{\percent} more effective in reaching the $ p $-spin ground state if a pause of length $ \lpause = \SI{900}{\nano\second} $ is inserted at $ s = \spauseopt $ during an annealing of duration $ \tf = \SI{100}{\nano\second} $, compared with a regular quantum annealing of time $ \tau' $ and no pause ($ \Phi \approx 0.957 $ versus $ \Phi \approx 0.664 $). In the isolated case, the fidelity is two orders of magnitude smaller ($ \Phi \approx \num{5.39e-2} $). We would need an annealing time $ \tf \approx \SI{50}{\micro\second} $ to achieve a fidelity similar to that at the optimal pausing point. These results are in qualitative agreement with those of Ref.~\cite{marshall}, and indicate that exploiting the environment in a controlled fashion allows to substantially reduce the time-to-solution compared with standard quantum annealing.

\section{Acknowledgments}

We acknowledge the CINECA award under the ISCRA initiative, for the availability of high performance computing resources and support. 
	    	    
\appendix	

\section{Validity of the Lindblad approximation}\label{app:lindblad}

The time-dependent Lindblad quantum master equation can be derived under very general assumptions~\cite{zanardi:master-equations}. The $ \text{qubit} + \text{bath} $ Hamiltonian reads $ \ham(t) = \ham_Q(t) + \ham_B + V\ped{QB} $, where all terms are defined in Section~\ref{sec:model}. The bosonic bath is at equilibrium at temperature $ T $ and is described by the correlation function $ \mathcal{B}(t) = \ev{X(t) X(0)} $, where $ X = \sum_{k}(a_k + a_k^\dagger) $ and $ X(t) $ is in the interaction picture with respect to $ \ham_B $. The modulus of $ \mathcal{B}(t) $ decays as $ \exp(-t/\tau_B) $ at small times and as $ (t/\tau_M)^{-2} $ at longer times, with $ \tau_B = \beta / 2\uppi + \bigO(\omegac^{-1}) $ and $ \tau_M = \sqrt{2\beta / \omegac} $~\cite{zanardi:master-equations}. 

In the weak coupling regime, the Born, Markov and rotating wave approximations lead to a quantum master equation of the form reported in Eq.~\eqref{eq:lindblad}. This is equivalent to disregarding correlations and memory effects, and to ensuring complete positivity.
The Lamb shift Hamiltonian and the dissipator super-operator in Eq.~\eqref{eq:lindblad} are expressed as follows:
\begin{subequations}\label{eq:hls-diss}
	\begin{gather}
		\ham\ped{LS} = \sum_{a \neq b} S(\omega_{ba}) L^\dagger_{ab} L_{ab} + S(0)\sum_{ab} L^\dagger_{aa} L_{bb},\\
		\begin{align}
			\diss\bigl[\rho(t)\bigr] &= \sum_{a \neq b} \gamma(\omega_{ba}) \Bigl(L_{ab} \rho(t) L^\dagger_{ab} - \frac{1}{2} \big\lbrace L^\dagger_{ab} L_{ab}, \rho(t)\big\rbrace\Bigr) \notag\\&\quad {}+ \gamma(0)\sum_{ab} \Bigl(L_{aa} \rho(t) L^\dagger_{bb} - \frac{1}{2} \left\lbrace L^\dagger_{aa} L_{bb}, \rho(t)\right\rbrace\Bigr),
		\end{align}
	\end{gather}
\end{subequations}
where 
\begin{equation}\label{eq:hilbert-transform}
	S(\omega) = \mathcal{P}\int_{-\infty}^{+\infty} \frac{d\omega'}{2\pi} \frac{\gamma(\omega')}{\omega - \omega'} ,
\end{equation}
$ \mathcal{P} $ denotes the principal value, and we omitted the time dependence from Lindblad operators and frequencies for brevity. We report here the constraints that all parameters must satisfy for the approximation to be valid. More details and extended discussion can be found in Ref.~\cite{zanardi:master-equations}.
\begin{itemize}
	\item The adiabatic condition requires $ \tf \gg h / \Delta^2 $, where $ h = \max_{s,a,b} \braket{\epsilon_a(s) | \partial_s \ham_Q(s) | \epsilon_b(s)} $. However, extensive use of adiabatic QMEs out of the adiabatic regime show results in reasonable agreement with other approaches. 
	\item The perturbative corrections to the unitary dynamics due to the dissipator super-operator $ \diss $ must be small compared with $ \Delta $. This implies $ g^2 \tau_B \ll \Delta $.
	\item The Markov approximation requires $ \lvert\mathcal{B}(t)\rvert $ to decay more rapidly than the typical relaxation time scale $ 1 / g $. Thus, $ g \tau_B \ll 1 $.
	\item The changes of the instantaneous eigenbasis of $ \ham_Q(t) $ must be small during a time scale of the order of $ \tau_B $. This requires $ \tf \gg h \tau_B^2 $.
\end{itemize}
Additionally, for Ohmic dissipation, $ \omegac $ must satisfy the following constraints:
\begin{gather}
	\beta \omegac \gg 1; \\
	\frac{1}{\omegac \log(\beta \omegac)} < \min\left\lbrace 2\tau_B, \frac{\tau_B h}{\tf} \left(\frac{1}{\Delta^2} + \frac{\tau_B^2}{\tf}\right)\right\rbrace.
\end{gather}
Except for the adiabatic condition, which we purposely violated as discussed in Section~\ref{sec:model}, all other conditions are satisfied by our parameters.

\section{Comparison with the search Hamiltonian}\label{app:search}

In Section~\ref{subsec:p-spin}, we discussed the similarity between the $ p $-spin system with $ \nspin = 20 $ and $ p = 19 $, and the typical search Hamiltonian of Eq.~\eqref{eq:search} (see Fig.~\ref{fig:hammings}). Indeed, these two systems can be mapped onto each other in the limit of large and odd $ p $. 
Here we will quickly review this mapping.

We work in the subspace with maximum total spin, \ie, $ S = \nspin / 2 $. Within this subspace we have $ \nspin + 1 $ wave functions identified by  $ \ket{w} \equiv \ket{\nspin - 2 w} $, with $ w = 0, 1, \dots, \nspin $. They are eigenstates of the $ z $-component of the total spin, $ \sum_i \sigma_i^z $, satisfying
\begin{equation}
	\biggl(\sum_i \sigma^z_i\biggr) \ket{w} = (\nspin - 2 w) \ket{w}.
\end{equation}
In this basis, Hamiltonian~\eqref{eq:pspin-hamiltonian} is represented as
\begin{equation}
	\ham\ped{p} = -\frac{\nspin}{2} \sum_{w = 0}^{\nspin} {(1 - 2 w / \nspin)}^p \ket{w}\bra{w}.
\end{equation}
Its ground state is the state $ \ket{0} $.
In the limit $ \nspin \to \infty $ and $ p \to \infty $ ($ p \leq \nspin $), Hamiltonians~\eqref{eq:pspin-hamiltonian} and~\eqref{eq:search} have all the same  eigenvalues for $ w = 0, 1, \dots, \nspin - 1 $, and differ only by their last eigenvalue. This state has very large energy and does not affect the outcome of the computation, with our parameters. In this sense, the Hamiltonian~\eqref{eq:pspin-hamiltonian} is analogous to the Hamiltonian~\eqref{eq:search}.

\begin{figure}[tb]
	\includegraphics[width = \linewidth]{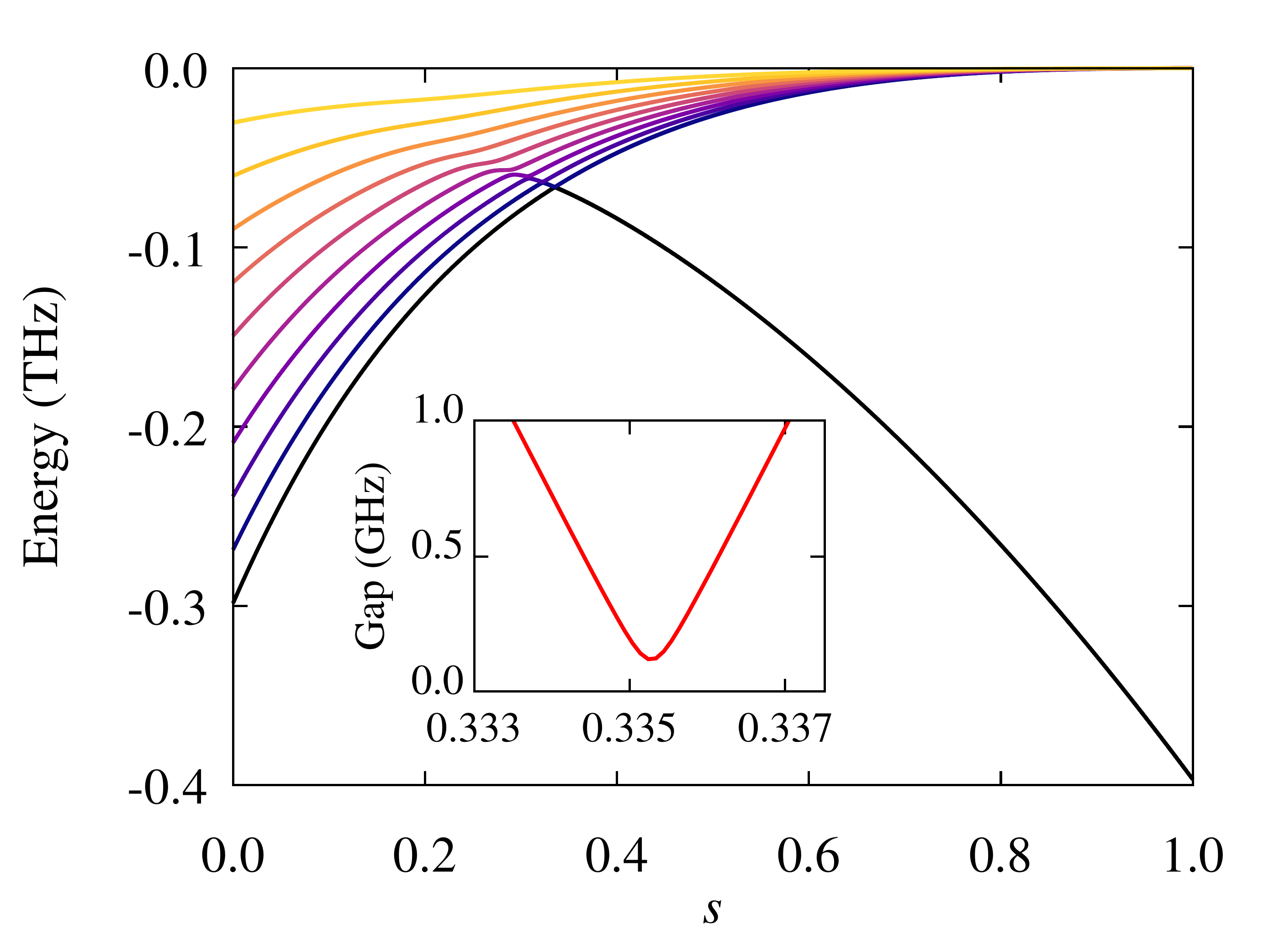}
	\caption{The $ L = 10 $ lowest-lying eigenvalues of the search Hamiltonian, for our choice of parameters. The inset shows the instantaneous gap between the ground state and the first excited state, around $ s = \sgap\api{search} \approx 0.335 $. Units are defined with $ \hslash = 1 $.}
	\label{fig:spectrum-search}
\end{figure}

In order to highlight the similarities between the $ p $-spin and the search Hamiltonian, we repeat part of the analysis presented in Section~\ref{sec:results}, using the search Hamiltonian.
In Fig.~\ref{fig:spectrum-search}, we report the first $ L = 10 $ eigenvalues of the Hamiltonian $ \ham_Q(s) $, as a function of the dimensionless annealing time $ s $. Comparing the eigenvalues of Fig.~\ref{fig:spectrum-search}  with the ones of the $ p $-spin model (Fig.~\ref{fig:spectrum}), we note a very similar behavior except for two minor differences. The first one is that, differently from the $ p $-spin case, in the search model all the excited energy levels at $ s = 1 $ have zero energy, and there are no anti-crossing involving pairs of excited eigenstates. The second difference is that the time of the minimal gap is slightly shifted ($ \sgap\api{search} \approx 0.335 $), and the minimal gap is slightly reduced ($ \Delta \approx \SI{0.12}{\giga\hertz} $). However, these differences do not affect the dynamics significantly.

\begin{figure*}[tb]
	\centering
	\subfloat[]{\label{fig:pause-search-length}\includegraphics[width = 0.49\linewidth]{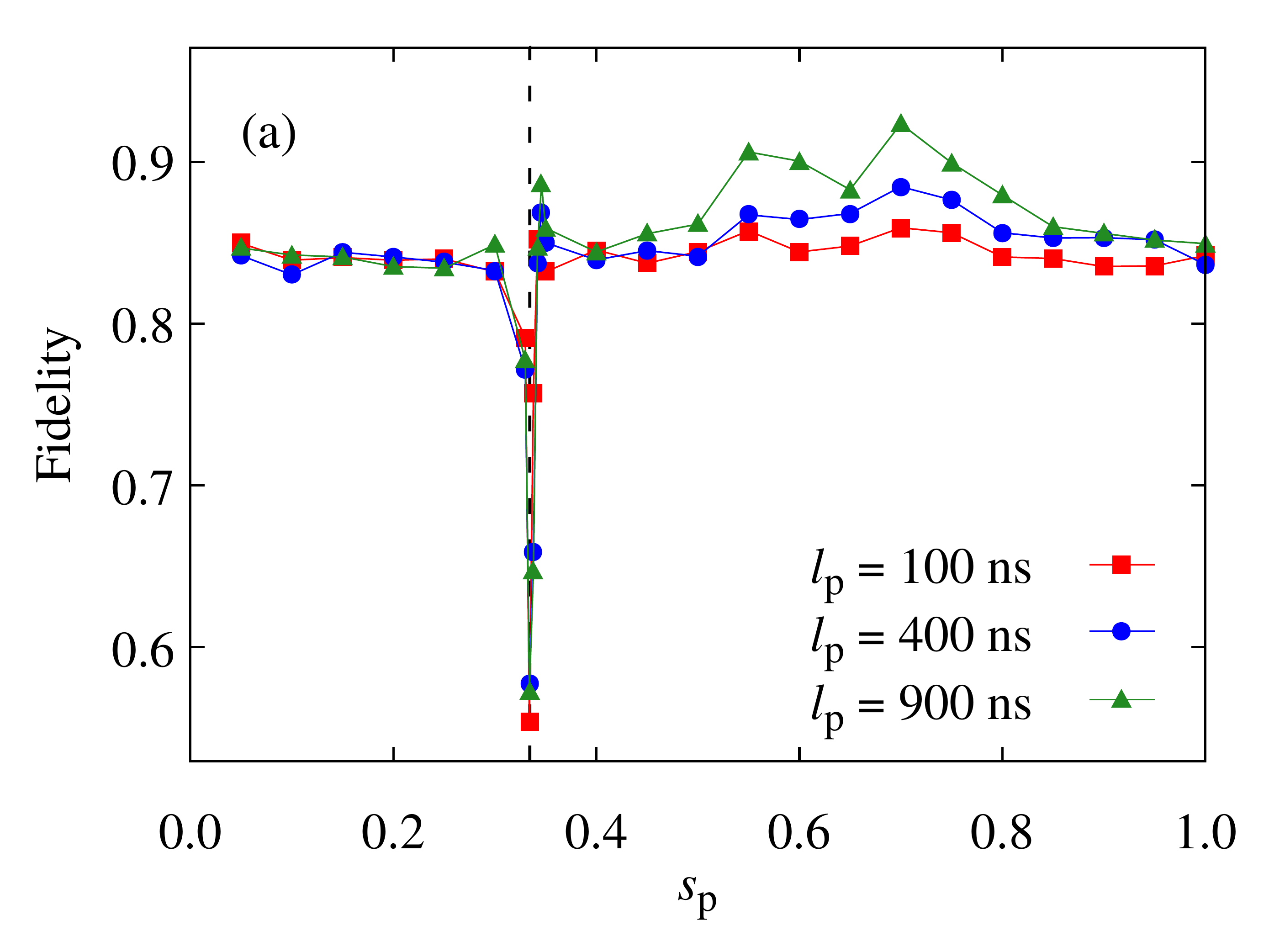}}\hfill
	\subfloat[]{\label{fig:pause-search-zoom}\includegraphics[width = 0.49\linewidth]{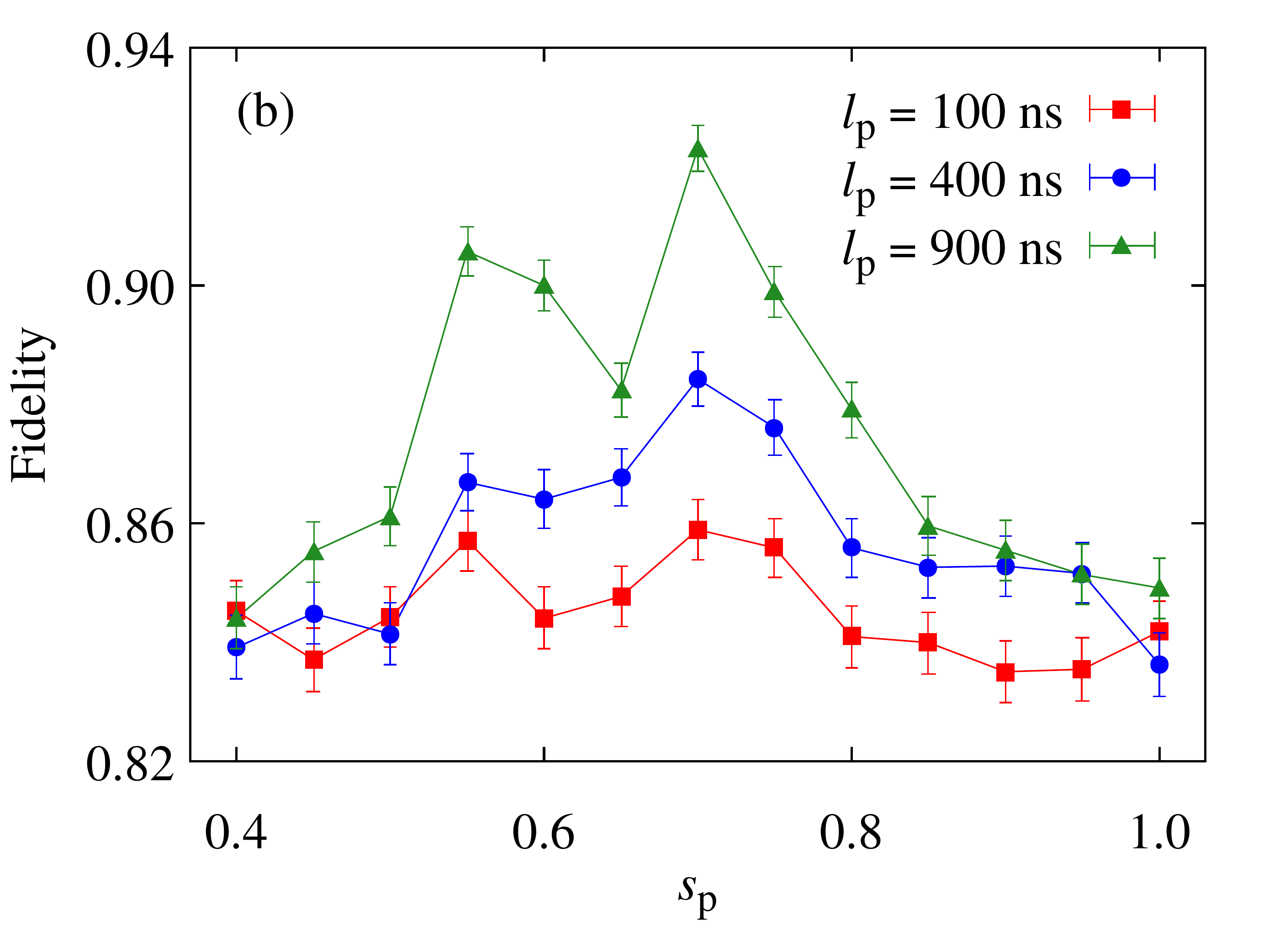}}
	\caption{Fidelity of the quantum annealing towards the search Hamiltonian of Eq.~\eqref{eq:search} as a function of the pausing time $ \spause $, for several values of the pause length $ \lpause $. Error bars are the standard errors of MCWF [see Eq.~\eqref{eq:mc-errors}]. In Fig.~\ref{fig:pause-search-length}, error bars are of the order of the point size, and the dashed vertical line indicates the time of the minimal gap, $ \sgap\api{search} \approx 0.335 $. In Fig.~\ref{fig:pause-search-zoom}, we focus on the region around the optimal pausing point $ \spauseopt\api{search} \approx 0.70 $.}
	\label{fig:pause-search}
\end{figure*}

Using an annealing time $ \tf = \SI{100}{\nano\second} $, we investigated the effects of pausing also using the search Hamiltonian as target, during a dissipative dynamics with the same parameters as those of Section~\ref{sec:results}. At $ s = 1 $, we measured the fidelity as a function of the pausing point $ \spause $ for several values of the pause duration $ \lpause $ (\eg, for $ \lpause = \text{\SIlist{100;400;900}{\nano\second}} $). Our results are reported in Fig.~\ref{fig:pause-search}. Panel~\ref{fig:pause-search-zoom} is a focus around the optimal pausing point, which in this case is $ \spauseopt\api{search} \approx 0.70 $.

Similarly to the $ p $-spin case, we distinguish four different regions.
\begin{enumerate}
	\item When $ \spause < \sgap\api{search} $, the level spacing is large and the dynamics is unaffected by pauses. The baseline fidelity is slightly larger than the $ p $-spin one ($ \Phi \approx 0.849 $ versus $ \Phi \approx 0.799 $).
	\item When $ \spause \approx \sgap\api{search} $, the mean level spacing is one order of magnitude smaller than the temperature. Thermal processes are frequent and deplete ($ \spause \lesssim \sgap\api{search} $) or repopulate ($ \spause \gtrsim \sgap\api{search} $) the adiabatic ground state.
	\item When $ \spause \approx \spauseopt\api{search} $, the fidelity increases. At difference with the $ p $-spin case, the behavior around the optimal pausing point is noisier and a secondary maximum appear for $ \lpause = \SI{900}{\nano\second} $.
	\item For $ \spause \approx 1 $, the fidelity goes back to its baseline.
\end{enumerate}
The qualitative behavior of these data is very similar to that of the $ p $-spin model (see Fig.~\ref{fig:pause}).

\begin{figure*}[tb]
	\centering
	\subfloat[]{\label{fig:populations-pspin}\includegraphics[width = 0.49\linewidth]{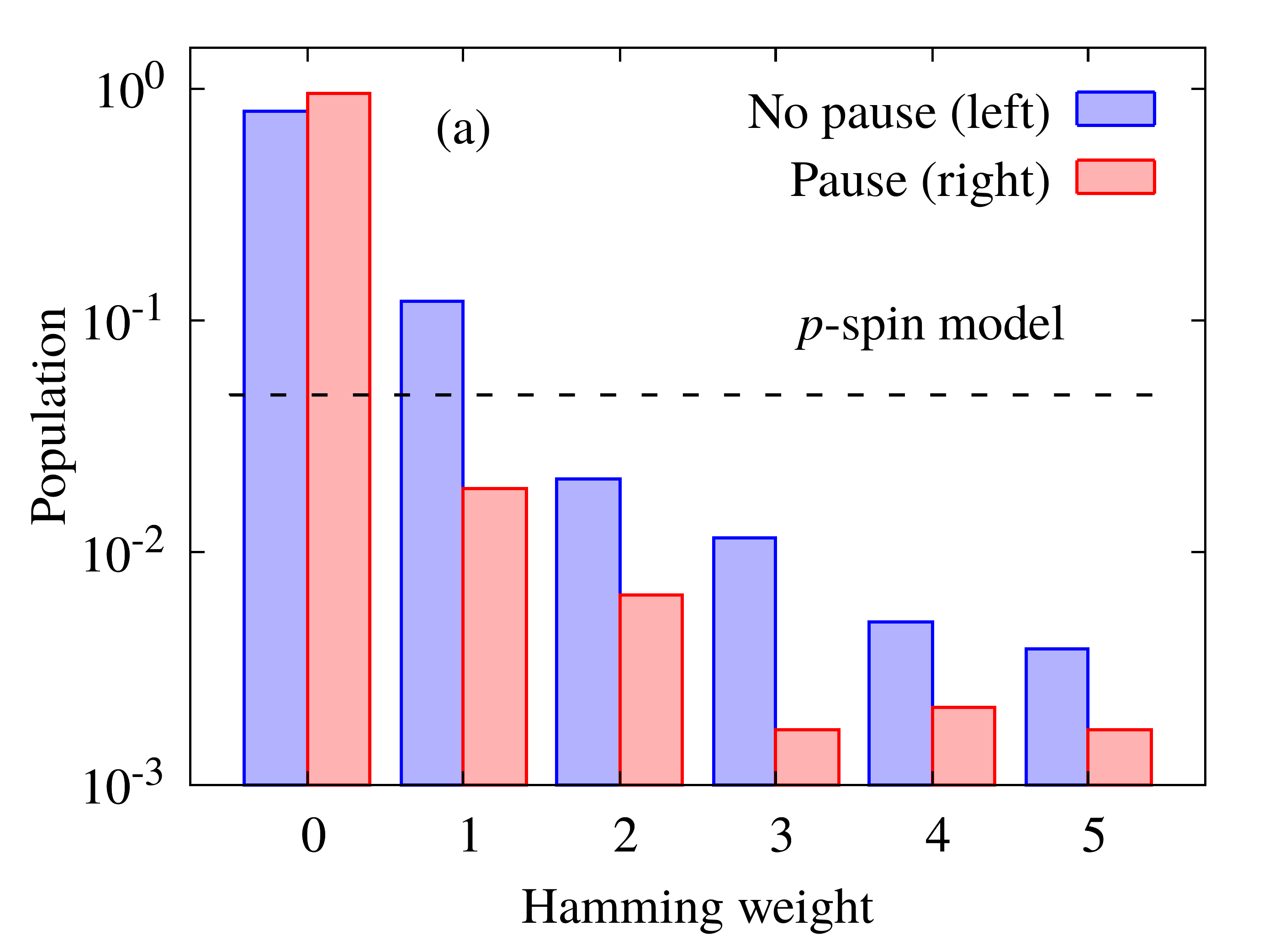}}\hfill
	\subfloat[]{\label{fig:populations-search}\includegraphics[width = 0.49\linewidth]{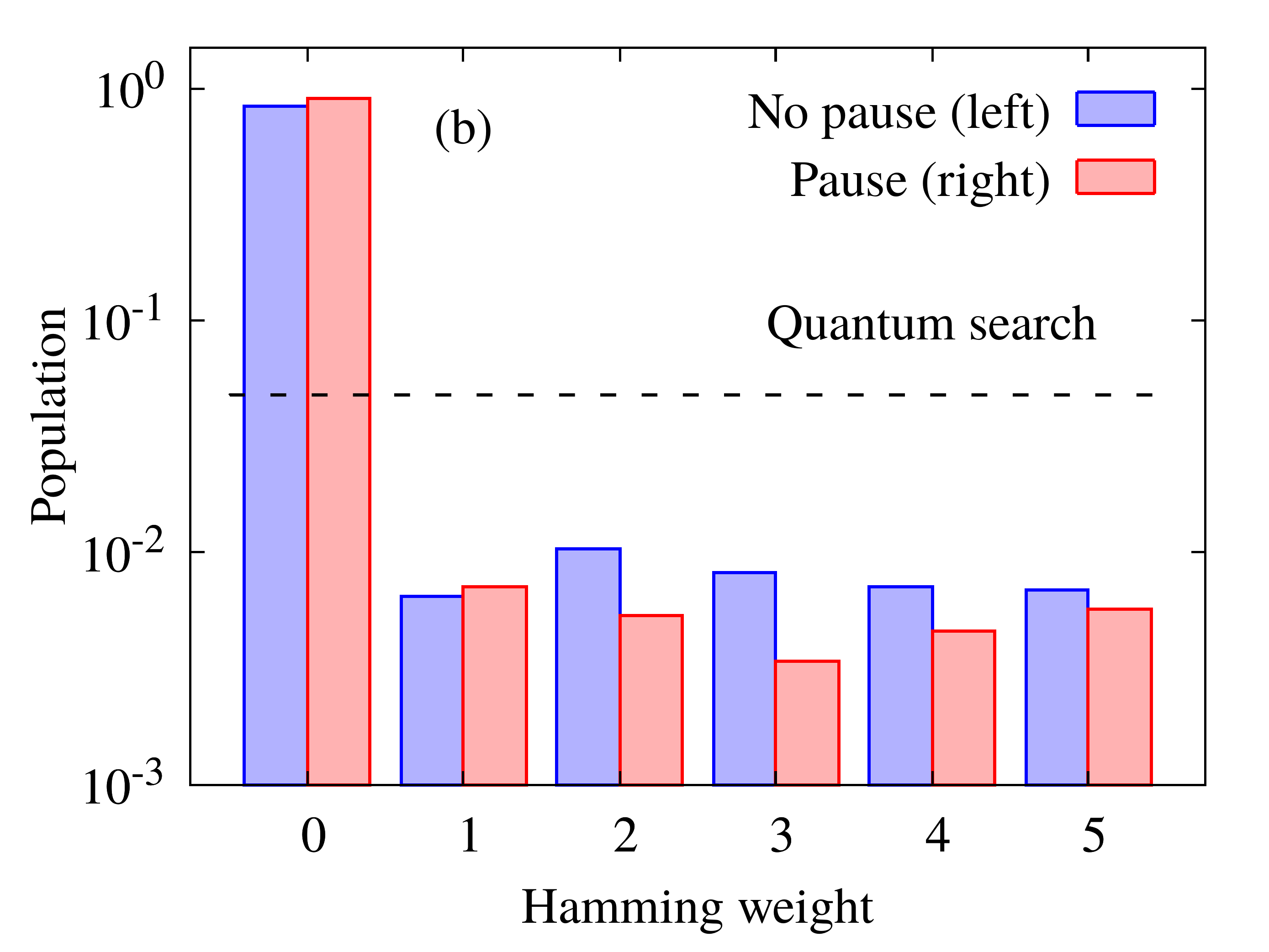}}
	\caption{Populations of the first $ L = 6 $ computational basis states as a function of the Hamming distance from the ferromagnetic ground state, at the end of a quantum annealing of duration $ \tf = \SI{100}{\nano\second} $. The left panel is for the $ p $-spin Hamiltonian, the right panel is for the search Hamiltonian. The blue bars on the left are data for a dynamics without pauses. The red bars on the right are data for a quantum annealing with a pause of $ \lpause = \SI{900}{\nano\second} $ inserted at the optimal pausing point ($ \spauseopt \approx 0.55 $ and $ \spauseopt\api{search} \approx 0.70 $, respectively). The black dashed line is $ P = 1/N $, with $ N = \nspin + 1 $.}
	\label{fig:populations}
\end{figure*}

We also computed all the diagonal elements of the qubit density matrix at the end of the annealing, for both the search Hamiltonian of Eq.~\eqref{eq:search} and the $ p $-spin Hamiltonian of Eq.~\eqref{eq:pspin-hamiltonian}, for the same starting state and all equal parameters. In Fig.~\ref{fig:populations}, we show the populations of the first $ L = 6 $ states, as a function of the Hamming distance from the ferromagnetic ground state, using a semilogarithmic scale. The left panel is for the $ p $-spin Hamiltonian and the right panel is for the search Hamiltonian. In both panels, and for each Hamming weight, the blue bars on the left refer to a dissipative quantum annealing with $ \tf = \SI{100}{\nano\second} $ and no pauses, whereas the red bars refer to a quantum annealing of the same duration, and a pause of $ \lpause = \SI{900}{\nano\second} $ inserted at the optimal pausing point ($ \spauseopt \approx 0.55 $ and $ \spauseopt\api{search} \approx 0.70 $, respectively). The black dashed line represents the (all equal) populations of the starting state, \ie, $ P = 1/N $, with $ N = \nspin + 1 $. 
The result shows striking similarities between the two cases.
However, some minor differences can be found. In particular, in the case of the $ p $-spin Hamiltonian the population of the first excited state, with Hamming weight $ w = 1 $,  is slightly larger with respect to the corresponding one using the search Hamiltonian. However, in both cases the final population of these states is at least one order of magnitude smaller than the ground state population. All other states are even less populated. Moreover, if a pause is inserted at the optimal pausing point, the $ p $-spin results become even closer to the ones of the search task, as the population of the state with $ w = 1 $ drops below $ P = 1/N $, and all other populations are about three orders of magnitude smaller than the ground state occupation probability.

\clearpage


%

\end{document}